\documentclass{aa}  

\usepackage{graphicx}
\usepackage{caption}
\usepackage{subcaption}
\usepackage{todonotes}
\usepackage{nicefrac}
\usepackage[switch]{lineno}
\usepackage{txfonts}
\usepackage{hyperref}
\usepackage{color, colortbl}
\usepackage{siunitx}
\usepackage{orcidlink}

\DeclareSIUnit\parsec{pc}
\DeclareSIUnit\mag{mag}
\DeclareSIUnit\days{days}

\defcitealias{Kenworthy2021SALT3}{K21}
\def\diff{}
\def\modelname{SALT3+}

\def\mainpaper{Rigault et al. (2024)}

\def\specpaper{Johannson et al. (2024)}

\def\colorpaper{Ginolin et al. (2024b)}
\begin{document} 

   \title{ZTF SN\, Ia DR2: Improved SN\,Ia colors through expanded dimensionality with SALT3+}

   \author{W. D. Kenworthy  \inst{1} \orcidlink{0000-0002-5153-5983}
\and A. Goobar \inst{1} \orcidlink{0000-0002-4163-4996}
\and D. O. Jones \inst{2} \orcidlink{0000-0002-6230-0151}
\and J. Johansson \inst{1} \orcidlink{0000-0001-5975-290X}
\and S. Thorp \inst{1} \orcidlink{0009-0005-6323-0457}
\and R. Kessler \inst{3,4} 
\and U. Burgaz \inst{5} \orcidlink{0000-0003-0126-3999}
\and S. Dhawan \inst{6} 
\and G. Dimitriadis \inst{5} \orcidlink{0000-0001-9494-179X}
\and L. Galbany \inst{7,8} \orcidlink{0000-0002-1296-6887}
\and M. Ginolin \inst{9} \orcidlink{0009-0004-5311-9301}
\and Y.-L. Kim \inst{10} \orcidlink{0000-0002-1031-0796}
\and K. Maguire \inst{5} \orcidlink{0000-0002-9770-3508}
\and T. E. Müller-Bravo \inst{9,10} \orcidlink{0000-0003-3939-7167}
\and P. Nugent \inst{11,12} \orcidlink{0000-0002-3389-0586}
\and J. Nordin \inst{13} \orcidlink{0000-0001-8342-6274}
\and B. Popovic \inst{9} \orcidlink{0000-0002-8012-6978}
\and P. J. Pessi \inst{1} \orcidlink{0000-0002-8041-8559}
\and M. Rigault \inst{9} \orcidlink{0000-0002-8121-2560}
\and P. Rosnet \inst{14} \orcidlink{0000-0002-6099-7565}
\and J. Sollerman \inst{15} \orcidlink{0000-0003-1546-6615}
\and J. H. Terwel \inst{5,16} \orcidlink{0000-0001-9834-3439}
\and A. Townsend \inst{12} \orcidlink{0000-0001-6343-3362}
\and R. R. Laher \inst{17} \orcidlink{0000-0003-2451-5482} 
\and J. Purdum \inst{18}  \orcidlink{0000-0003-1227-3738}
\and D. Rosselli \inst{19}\orcidlink{0000-0001-6839-1421}
\and B. Rusholme\inst{17} \orcidlink{0000-0001-7648-4142} 
}
          
   \institute{The Oskar Klein Centre, Department of Physics, Stockholm University, SE - 106 91 Stockholm, Sweden
          \\ \email{darcy.kenworthy@fysik.su.se}
\and Institute for Astronomy, University of Hawai‘i, 640 N.\ Aohoku Pl., Hilo, HI 96720, USA
\and
Kavli Institute for Cosmological Physics, University of Chicago, Chicago, IL 60637, USA
\and
Department of Astronomy and Astrophysics, University of Chicago, Chicago, IL 60637, USA
         \and
         School of Physics, Trinity College Dublin, The University of Dublin, Dublin 2, Ireland
\and
Institute of Astronomy, Madingley Rd, Cambridge CB3 0HA, United Kingdom
\and
              Institute of Space Sciences (ICE, CSIC), Campus UAB, Carrer de Can Magrans, s/n, E-08193 Barcelona, Spain 
             \and Institut d'Estudis Espacials de Catalunya (IEEC), 08860 Castelldefels (Barcelona), Spain 
\and
Univ Lyon, Univ Claude Bernard Lyon 1, CNRS, IP2I Lyon/IN2P3, UMR 5822, F-69622, Villeurbanne, France
\and Department of Physics, Lancaster University, Lancs LA1 4YB, UK
\and
Lawrence Berkeley National Laboratory, 1 Cyclotron Road MS 50B-4206, Berkeley, CA, 94720, USA
\and
Department of Astronomy, University of California, Berkeley, 501 Campbell Hall, Berkeley, CA 94720, USA
\and 
Institut für Physik, Humboldt-Universität zu Berlin, Newtonstr. 15, 12489 Berlin, Germany
\and
Université Clermont Auvergne, CNRS/IN2P3, LPCA, F-63000 Clermont-Ferrand, France
\and
The Oskar Klein Centre, Department of Astronomy, Stockholm University, SE -106 91, Stockholm, Sweden
\and
Nordic Optical Telescope, Rambla José Ana Fernández Pérez 7, ES-38711 Breña Baja, Spain
\and Caltech Optical Observatories, California Institute of Technology, Pasadena, CA 91125, USA
\and  Aix Marseille Université, CNRS/IN2P3, CPPM, Marseille, France }

   \date{Received September 15, 1996; accepted March 16, 1997}

 
  \abstract
   {Type Ia supernovae (SNe Ia) are a key probe in modern cosmology, as they can be used to measure luminosity distances at gigaparsec scales. Models of their light-curves are used to project heterogeneous observed data onto a common basis for analysis. }
   {The SALT model currently used for SN Ia cosmology describes SNe as having two sources of variability, accounted for by a color parameter $c$, and a ``stretch'' parameter $x_1$. We extend the model to include an additional parameter we label $x_2$, to investigate the cosmological impact of currently unaddressed light-curve variability.}
   {We construct a new SALT model, which we dub ``\modelname''. This model was trained by an improved version of the \texttt{SALTshaker} code, using training data combining a selection of the second data release of cosmological SNe\,Ia from the Zwicky Transient Facility and the existing SALT3 training compilation. }
   { We find additional, coherent variability in supernova light-curves beyond SALT3. Most of this variation can be described as phase-dependent variation in $g-r$ and $r-i$ color curves, correlated with a boost in the height of the secondary maximum in $i$-band. These behaviors correlate with spectral differences, particularly in line velocity. We find that fits with the existing SALT3 model tend to address this excess variation with the color parameter, leading to less informative measurements of supernova color. We find that neglecting the new parameter in light-curve fits leads to a trend in Hubble residuals with $x_2$ of $0.039 \pm 0.005$ mag, representing a potential systematic uncertainty. However, we find  no evidence of a bias in current cosmological measurements. }
   {We conclude that extended SN Ia light-curve models promise mild improvement in the accuracy of color measurements, and corresponding cosmological precision. However, models with more parameters are unlikely to substantially affect current cosmological results.}


   \maketitle
%

\section{Introduction}
\label{sec:intro}

Type Ia supernovae are standardizable candles which allow precision measurements of cosmological distances across cosmic history at gigaparsec scales. As a homogeneous population of objects sourced from the thermonuclear explosion of a white dwarf, they show extraordinary spectroscopic and photometric consistency. Measurements of Type Ia supernovae (SNe\,Ia) from photometric surveys are a key ingredient of modern efforts to measure the properties of dark energy \citep{TheLSSTDarkEnergyScienceCollaboration2018,Hounsell2018,Brout2022PantheonPlus,Rubin2023UNITY,Vincenzi2024DESCosmo}. Their use as cosmological indicators relies on the use of a light-curve model serving several purposes: a) to interpolate samples from multiple surveys onto a common phase and wavelength basis for comparison, b) reduce the dimensionality of the data, and c) parameterize the variability of the underlying data so they can be used by standardization and cosmology analysis.
 
The Spectral Adaptive Light-curve Template (SALT) model was first published in \citet{Guy2005}, {\diff and was substantially updated in \citet{Guy2007}}. With ongoing improvements to sample size and calibration, SALT2 was the primary model for every published measurement of the dark-energy equation-of-state parameter $w$ using SNe\,Ia between 2011 and 2022. The model was updated in \cite{Guy2010} and \citet{Betoule2014}. A more recent revision (SALT2.T21) was presented by \citet{Taylor2021SALT2}. In \citet{Kenworthy2021SALT3} (hereafter \citetalias{Kenworthy2021SALT3}) a new training code was developed, the model error term was redefined, and the training sample was expanded by a factor of 2.5, resulting in the SALT3 model (here referred to as SALT3.K21). Further work from \citet{Dai2023Propagating} and \citet{Taylor2023SALT2vsSALT3} has focused on constraining systematics of the \texttt{SALTShaker} training code and SALT3.K21 model, or incorporating correlations with host galaxy properties \citep{Jones2023, Taylor2024}. The Union 3/UNITY 1.5 analysis  \citep{Rubin2023UNITY}, as well as the  5 year analysis of the Dark Energy Survey's supernova experiment \citep{Taylor2023SALT2vsSALT3,Vincenzi2024DESCosmo}, used SALT3 for the first time in a cosmological analysis. \cite{Jones2023} also explored variation in light curves between different host-galaxy types using the SALT framework.

The SALT framework is distinguished by several key elements from other SN\,Ia models available in the community (e.g. SnooPy \citep{Burns2014IntrinsicSupernovae,Burns2018CSPHubble}, BayeSN \citep{Mandel2011,Thorp2021TestingDr1,Grayling2024BayeSNScalable}, MLCS2k2 \citep{Jha2007}, SNEMO \citep{Saunders2018}, SUGAR \citep{Leget2019}). The model is fully empirical, by construction prioritizing the parametrization of SN\,Ia photometric diversity over links to specific physical mechanisms e.g.\ spectroscopic lines, theoretical simulations, or specific parametrizations of dust extinction. Spectroscopic data is included in model construction to ensure rest-frame modelled photometry is reliable across the redshift range and prevent deconvolution noise\footnote{When inferring spectra from photometry (a convolved SED), narrow oscillations in the reconstruction are a common issue due to the lack of spectral resolution. This phenomenon is known as deconvolution noise.}, but is explicitly downweighted compared to photometric data. SALT aims to avoid making assumptions about cosmology or underlying populations in model construction. Instead SALT prioritizes homogenizing and reducing diverse measured photometry to a form suitable for further analysis by frameworks such as Bayesian Estimation Applied to Multiple Species with Bias Correction (BEAMS/BBC) \citep{Kunz2007BEAMS,KesslerScolnic2017}, Unified Nonlinear Inference for Type-Ia cosmologY (UNITY) \citep{Rubin2015UNITY,Rubin2023UNITY}, Dust2Dust \citep{Popovic2021Dust2Dust}, and others \citep[e.g.][]{March2011, Shariff2016, Mandel2017, Feeney2018, Rahman2022, Wojtak2023TwoPop}. These frameworks are better suited to make and evaluate decisions about selection, cosmology, and population demographics. Finally, as an empirical model, SALT is constructed by reference to a training sample of SN\,Ia data. The construction of this sample has emphasized the use of data from a diversity of carefully collected SN\,Ia surveys. By doing so, we aim to include demographics similar to samples typically used for cosmology, across a range of filters, cadences, calibrations, and redshifts. By sample-matching in this way, both ``known unknown'' and ``unknown unknown'' systematics can be reduced.

\begin{figure}
    \centering
    \begin{subfigure}[t]{0.40\textwidth}
        \centering
        \includegraphics[width=\linewidth]{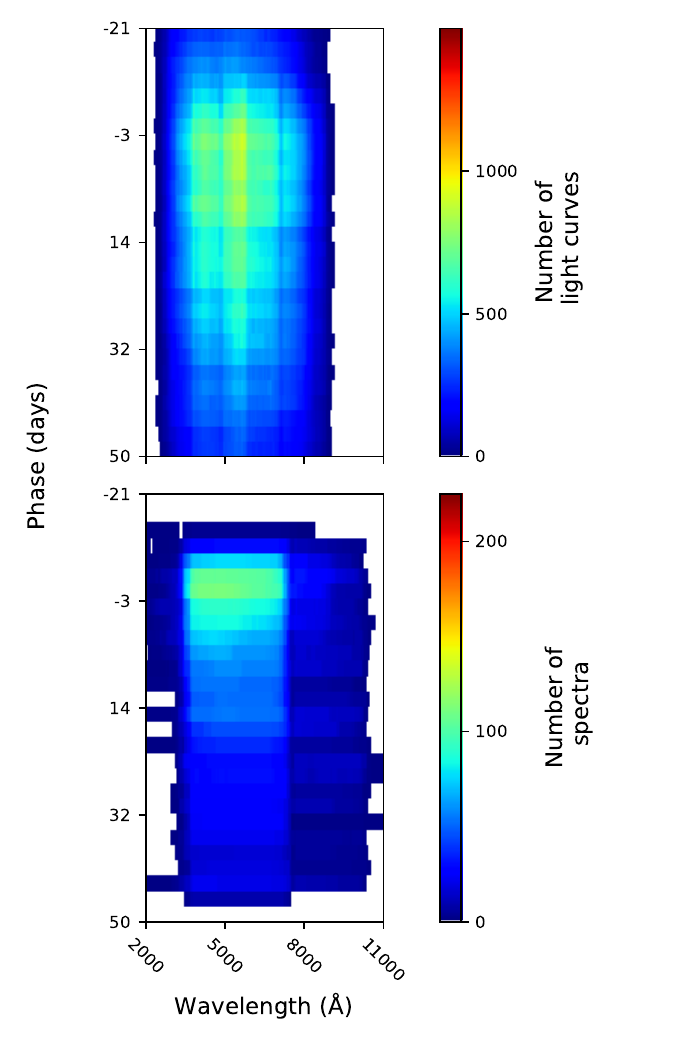} 
        \caption{Density of data with \citetalias{Kenworthy2021SALT3} compilation alone} 
    \end{subfigure}
    \vfill
    \begin{subfigure}[t]{0.40\textwidth}
        \centering
        \includegraphics[width=\linewidth]{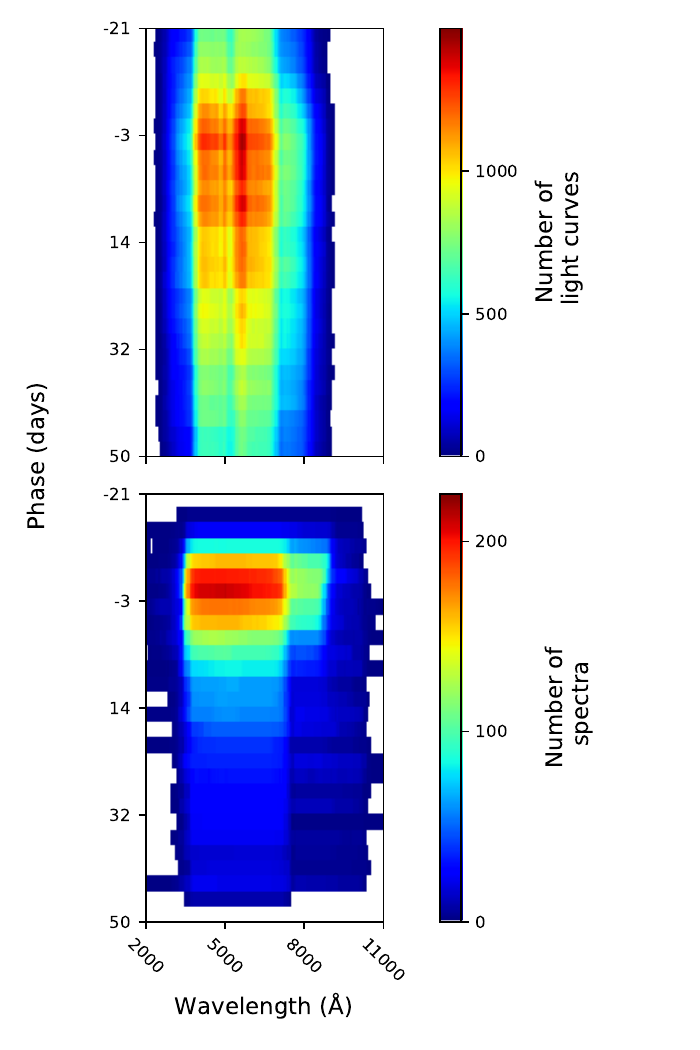} 
        \caption{Density of data with the addition of ZTF data to the \citetalias{Kenworthy2021SALT3} compilation}
    \end{subfigure}
    \caption{Density of spectroscopic and photometric data from training samples as a function of wavelength. A photometric epoch is considered to cover a given wavelength bin if the wavelength is within the FWHM of the filter in the rest-frame. Bin sizes are dictated by the underlying resolution of the SED model.} \label{fig:datacomparison}
\end{figure}

SALT models as published rely on a single parameter for the phase-dependent variation of the SN\,Ia light curve and a single color parameter accounting for intrinsic and/or  extrinsic reddening. The intrinsic  variation parameter $x_1$ is associated with the light-curve ``stretch'', a change in the decay time of the light-curve post-maximum. The stretch of SNe\,Ia has been associated with the amount of $^{56}$Ni produced in the thermonuclear explosion \citep{Phillips1993,Kasen2007WidthLuminosity}. The decay of this unstable isotope heats the ejecta, increasing the luminosity of the explosion as well as delaying ionization transitions of iron-group elements, {\diff increasing opacity and} broadening the light-curve. However theoretical evidence suggests that other factors may also affect the light-curve of the supernova. 

Observationally, work by \citet{Saunders2018} and \citet{Rubin2020Dimensionality} suggest that SN\,Ia phase-dependent spectrophotometry is more complicated than can be modeled by a single intrinsic parameter. Rigault et al. (2024) shows deviations of light-curve residuals from SALT fits in the ZTF DR2 sample, which suggests that the time-dependence of SN Ia colors are not fully addressed by SALT. \citet{Hayden2019StandardizationRise} split the SALT2 model into pre- and post-maximum components (extending the dimensionality by one). They found that the pre-maximum light-curves of SNe\,Ia seemed to be more effective in standardization than post-maximum. Fully trained SN Ia models, including SNEMO \citep{Saunders2018} and SUGAR \citep{Leget2019} have included higher-dimensionality variants. However both these models were exclusively developed with spectrophotometric data from the SNFactory \citep{Aldering2002SNFactory}. This unique data gave these models excellent coverage over the wavelength domain of the SNIFS instrument ($\sim 3300-8600$~\AA). However use of these models for cosmology analysis is obstructed by two primary issues. \citet{Rose2020SNEMOEval} found that the  majority of available SN\,Ia photometry was unable to constrain the 7-dimensional SNEMO7 model due to the size of the parameter space, with substantial correlations in fitted parameters which obstruct cosmological analysis. Further, the low redshift of the SNFactory training sample means that the model is possibly susceptible to evolving demographic biases with redshift. \citet{Rose2020SNEMOEval} therefore suggested that a lower-dimensionality model would likely be most useful. Therefore, it is desirable to make available a SN Ia model that incorporates the broad demographics and wavelength coverage of the SALT training samples, with a lower dimensionality than SNEMO7 to allow more robust fits against multi-band SN\,Ia photometry of typical observation cadence. Additionally, a higher dimensional BayeSN model was also included in the appendix of \citep{Mandel2022}. Beyond any utility in standardization,  a higher-dimensional SALT model may be important for use in bias correction simulations using BBC, or equivalently, selection integrals in UNITY.

Here we use the \texttt{SALTshaker} code presented in \citetalias{Kenworthy2021SALT3} to train a model we label \modelname, a two-dimensional model of phase-dependent SN Ia variability. To efficiently constrain phase-dependent behavior, we incorporate data from the Zwicky Transient Facility \citep{bellm2019,graham2019} in concert with the compilation of training data already assembled, which we discuss in Sec. \ref{sec:data}. We present updates to the \texttt{SALTshaker} code as well as the full description of the \modelname\ model in Sec. \ref{sec:SALTshaker}.  Lastly, we discuss our results and conclusions in Sec. \ref{sec:results} and \ref{sec:conclusions}.

\section{Data}
\label{sec:data}
We make use of the original compilation of data assembled in \citetalias{Kenworthy2021SALT3}, and add spectroscopically typed SNe\,Ia from the second data release of ZTF SNe\,Ia, ZTF SN Ia DR2 (Rigault et al. 2024, Smith et. al. 2024) to train our light-curve model. Combining these data increases the total number of photometric epochs and spectra by $\sim 50\%$. In Fig. \ref{fig:datacomparison}, we show the available data as a function of phase and wavelength for \citetalias{Kenworthy2021SALT3} as well as the ZTF sample. As we examine host galaxy relations in our analysis in Sect. \ref{subsec:cosmo}, we use host-galaxy masses from Pantheon+ \citep{Brout2022PantheonPlus} for the K21 compilation where available, and estimates from the ZTF DR2 for ZTF objects (Smith et al. 2024).

\subsection{K21 Compilation}

The K21 compilation was described in \citet{Kenworthy2021SALT3}. The compilation builds on the sample used by the Joint Light-curve Analysis to train the SALT2 model \citep{Betoule2014}, and incorporates data from many of the largest available surveys for cosmological SNe. In total, it consists of 1048 SNe from the Sloan Digital Sky Survey (SDSS; \citealp{Holtzman08,Kessler2009,Sako2018}), the Supernova Legacy Survey (SNLS; \citealp{Astier2006}, with spectra from \citealp{Walker2011,Balland2018} and private communication with M.\ Betoule, C.\ Balland), the Calan-Tololo Survey \citep{Hamuy1996}, the Center for Astrophysics surveys \citep[CfA;][]{Riess1999,Jha2006,Hicken2009a,Hicken2012}, the Carnegie Supernova Project \citep{Krisciunas2017}, the Foundation Supernova Survey \citep{Foley2018, Jones2019}, the Pan-STARRS Medium Deep Survey \citep{Rest2014, Scolnic2018}, and the Dark Energy Survey \citep{Abbott2019}. This sample was used to train the first published SALT3 models. Since then new calibration solutions for the photometric systems have been made available in \citet{Brout2022Fragilistic}\footnote{These calibration solutions were also used by \citep{Taylor2023SALT2vsSALT3} to retrain SALT3.}, and the models trained here have made use of these recalibrations. 

Following the suggestions of \citet{Vincenzi2024DESCosmo} as well as \citet{Taylor2023SALT2vsSALT3}, we exclude all\textit{ U}-band light curves from the sample due to calibration uncertainties (see also earlier discussion in \citealp{Kessler2009, Krisciunas2013}). This change affects 97 SNe in the data, principally from the CfA surveys.

\subsection{ZTF DR2}

As described in \mainpaper, the second data release of cosmological SNe from the ZTF collaboration  \citep{bellm2019,graham2019,Masci2019Data,Dekany2020ZTFObserving} consists of 3627 spectroscopically classified SNe Ia primarily obtained from the Bright Transient Survey\citep{fremling2020,perley2020}, a magnitude-limited sample of extragalactic transients in the northern sky. All SNe in this data release were observed between April 2018 and December 2020. We make use of the $gri$ band photometry from \mainpaper{} to train our models. The calibration solution of the ZTF SN sample is still a work in progress, and the models presented here will thus have unbudgeted calibration uncertainties of > 0.01 mag.

To ensure each individual ZTF light-curve used had sufficient coverage to examine the time-dependent behavior beyond $x_1$, we included substantial light-curve quality cuts. These cuts on ZTF data are stricter than those applied to the \citetalias{Kenworthy2021SALT3} sample, and we have chosen not to apply them to the previously compiled data. The primary reason for this is that the SALT model philosophy, as discussed in Sec. \ref{sec:intro}, prioritizes carefully collected yet \textit{heterogeneous} data. Applying these stricter cuts to the previous training sample would substantially reduce the amount of non-ZTF data available in the training, causing the model to prioritize explaining the ZTF data. {\diff Analogy here with principal component analysis is useful. When a PCA is performed, the vectors found depend on the importance of different modes in the data. As that distribution shifts, the relative importance of different modes shifts as well. Phenomenology is then allocated to different component vectors. Further discussion of the issues of imbalanced data-sets can be found in \citet{Fernandez2018}}. The overall ZTF sample includes $\sim 1600$ SNe passing the K21 quality cuts with host-redshift data, larger than the entirety of the previous training sample.  Including this amount of data naively would result in a model strongly weighted towards characterizing the ZTF data. Applying such a model to high-redshift SNe, measured in different bands and with possibly divergent demographics, might fail to capture relevant information. While these effects can be simulated using validation pipelines such as that presented in \citet{Dai2023Propagating}, preventing the ZTF data from becoming a majority of measured light-curves is a safer solution.  Accordingly we require, labeling phase relative to SALT2 maximum light as $p$ {\diff and binning photometric measurements falling within one day of each other in the same band}:

\begin{itemize}
    \item At least one measurement in at least two bands after peak brightness ($5 < p < 20$), to constrain the shape and color.
    \item At least five {\diff epochs in any bands} between $-20 < p <-1$ to ensure coverage of the rising light-curve.
    \item At least one epochs in all three $gri$ bands between $-10 < p < 35$.
    \item Classified as a normal SN\,Ia, or belonging to the 91T subtype.
    \item Rejected any supernova whose only available redshift had been measured by \texttt{snid}.
\end{itemize}

530 SNe from the ZTF data release pass these cuts and are included in our training sample. In addition to their photometric data, we include spectra, further detailed in \specpaper. This data is primarily from the SEDm spectrograph \citep{blagorodnova2018,rigault2019,Kim2022SEDM}, although contributions from other instruments and sources are present. All spectra from instruments other than SEDm are preprocessed using tools from kaepora \citep{Siebert19} to clip host-galaxy lines and estimate uncertainties. Previously, these tools were used in the construction of the \citetalias{Kenworthy2021SALT3} compilation.  For SEDm spectra, we rely on the pySEDm infrastructure \citep{rigault2019}. 

As can be seen from Fig. \ref{fig:datacomparison}, inclusion of ZTF data leads to a significant increase in the overall compilation of data. Unfortunately, many of the spectra contribute to the sample only in the portions of the phase and wavelength space that are already covered by existing data, particularly in the region immediately before maximum light. The photometric data are a sample with excellent phase coverage. The ZTF survey cadence results in many more SNe discovered at early phases. Further, increased sampling increases the reliability of the more flexible, higher dimensional model, reducing the amount of implicit interpolation performed by the fitter.

\subsubsection{Uncertainty Estimates}
\label{subsec:errors}
\begin{figure}
    \centering
    \includegraphics{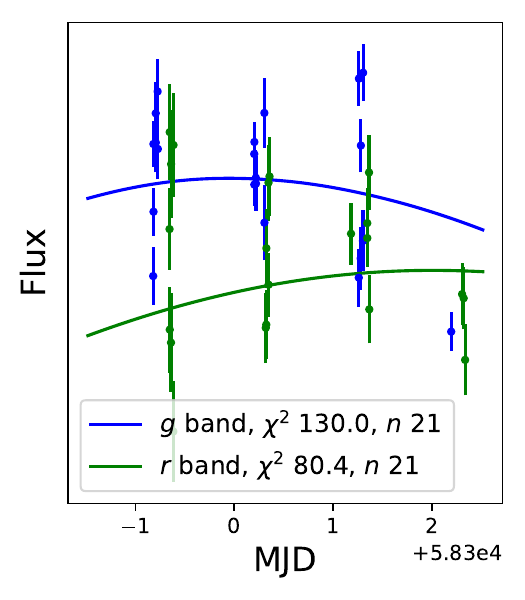}
    \caption{Light-curve of 2018cvq, shown at maximum light, along with the modeled SALT2 light-curve, as well as $\chi^2$ relative to raw photometric uncertainties. 2018cvq was selected as the SN\,Ia in the sample with the largest number of epochs at maximum light, where evolution of the light-curve is smallest.}
    \label{fig:photdispersion}
\end{figure}

\begin{figure*}
    \centering
    \includegraphics{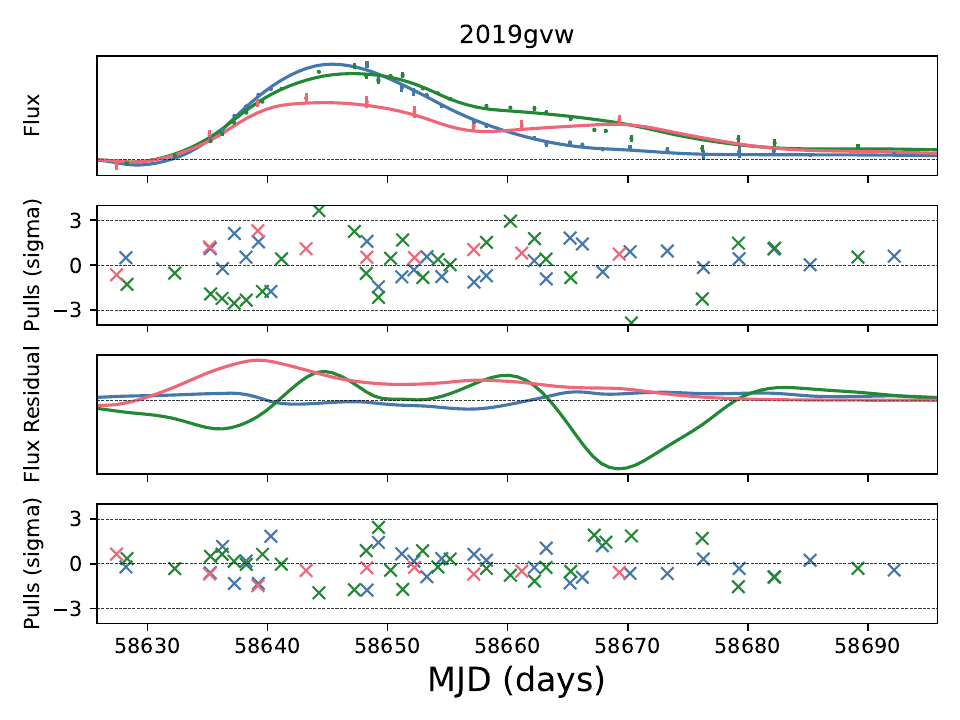}
    \caption{Illustration of the Gaussian process inference described in Section \ref{subsec:errors} as applied to the light-curve of 2019gvw. First panel: SALT2 fit to the light-curve, along with observations (binned to 1 day resolution). Second panel: residuals of SALT2 fit. These residuals show both correlated behavior and uncorrelated noise. Third panel: Gaussian process, conditioned on observed residuals, with hyperparameters from Table \ref{tab:photerrors}, showing the possible correlated structure we ultimately hope to include in the trained SALT model. Fourth panel: leave-one-out residuals from Gaussian process, showing expected scatter and reduced correlation.  }
    \label{fig:photerrors}
\end{figure*}

Light-curve model training is particularly sensitive to uncertainty estimates, as compared to fitting an existing light-curve model. For light-curve fitting, an individual point noisier than expected in a well-sampled light-curve represents a small fraction of the data, and the fit has only a few free parameters which will likely be dominated by the rest of the data. For model training however, there are thousands of underlying parameters determined by the training code. Individual points have a much stronger effect on parameters controlling the local region of phase/wavelength space which they cover, and can bias parameter inference for other SNe measured in that same region. We therefore evaluate the uncertainty estimates of the ZTF DR2 for their suitability for model training. 

Original error estimates from the ZTF data release, derived from difference imaging, are smaller than is sufficient to explain variation in the data. For example,  photometric measurements of 2018cvq shown in Fig. \ref{fig:photdispersion} taken within a day of peak show dispersion about the mean of 0.043 mag in $g$-band, while error estimates predict a dispersion of 0.018 mag. It is unlikely that this variance is explained by unmodelled but physical variation, as many of these observations are within the same night, and we conclude that the uncertainties are underestimated. Smith et al. (2024) suggests the use of error floors for the data, calculated based on residuals to the SALT2 model, of 0.025 mag, 0.035 mag, and 0.06 mag for $gri$ bands respectively. 

However an approach attributing outliers from light-curve fits to unbudgeted errors would potentially wash out exactly the light-curve features we want to examine in this work. In order to mitigate this issue as best as possible, we assume that the photometry errors are uncorrelated in time, while physical light-curve features will show correlation on a timescale of $\sim 5$ days. We then fit light-curve residuals (relative to the SALT2 model with parameters published in the main data release) with a Gaussian process to determine the amplitude of each contribution. 

We use the package celerite2 \citep{celerite1,celerite2} to fit the sample of light-curve data. For a given light-curve in some band $X$, we define $\sigma_X$ as the estimate of the relative unbudgeted, uncorrelated photometric error. We define the nuisance parameter $\sigma_X^\text{Corr}$ as the relative amplitude of correlated, potentially physical, light-curve variation in percentage units. We take the mean of the process to be the predicted flux of the fiducial SALT2 fit from the DR2 sample $F_X(p)$, where $p$ is the phase of the observation, and the kernel of the Gaussian process has two components: a) a diagonal, uncorrelated error floor equal to the sum in quadrature of the budgeted photometric error and $\sigma_X $ in magnitude units, and b) a Matérn 3/2 kernel with 5 day scale length and amplitude equal to $\sigma_X^\text{Corr}$ in magnitude units. The hyperparameters $\sigma_X^\text{Corr}$ and $\sigma_X$ are taken to be shared across the ZTF sample.  We make no K-corrections for this simple model of unbudgeted errors, assuming that the restricted redshift range of the ZTF sample will mean that such effects are small in the residuals to the SALT2 model. With three bands, we have six parameters to determine. We fit these Gaussian processes simultaneously across the whole sample, then maximize the log-likelihood of the fit with respect to the six parameters. Our fitted parameters are shown in Table \ref{tab:photerrors}, and an example light-curve fit is presented in Figure \ref{fig:photerrors}. 

We find that coherent variations in the residuals are detected in all 3 bands, particularly in $i$-band, confirming that there are likely variations in SN Ia light-curves beyond those addressed by the SALT2 model. We also find significant uncorrelated errors at scales $\sim 1-2\%$.

\begin{table}[]
    \centering
    \begin{tabular}{c|cc}
       Photometric Filter & $\sigma_X^\text{Corr}$ & $\sigma_X$ \\
        &  (centimag) & (centimag)  \\
        \hline
       ZTF \textit{g} & 4.6 & 1.3\\\
       ZTF \textit{r} & 4.9 & 1.8 \\
       ZTF \textit{i} & 8.4 & 2.2 
    \end{tabular}
    \caption{Estimated unbudgeted uncertainties in the ZTF photometric dataset. In each band, we assume some portion of the residuals from SALT2 light-curve fits is unbudgeted photometric error ($\sigma_X$). The rest of the variation is assumed to be physical variation in the light-curves, and correlated at a timescale of 5 days ($\sigma_X^\text{Corr}$).  }
    \label{tab:photerrors}
\end{table}

Based on these results we add (in quadrature) an error floor  to the estimated photometric uncertainties equal to $\sigma_X \cdot F_X(p)$ to all photometry incorporated into the training data. However we note that leave-one-out testing of the Gaussian process fits finds $3\sigma$ outliers at rates approximately 5 times higher than predicted by this model (756 outliers vs 156 predicted), implying that our error floors may not capture non-Gaussian noise in the data. Further, the Gaussian process used here cannot correct distinguish time-correlated photometric errors (effects from correlation in seeing, bad subtractions, etc.) from variation in the true light-curve. And as is detailed in \mainpaper and Lacroix et al. 2024 the calibration of the ZTF data is ongoing work. As a result we are unable to conclude that our model presented in this work will be of quality sufficient for use in cosmological analysis, and cannot recommend use in that context until we have reached a better understanding of statistical and calibration uncertainties in the ZTF data. Future photometric releases are expected to use a scene-modelling pipeline, and are expected to greatly improve the consistency of uncertainty estimates.

\section{SALT and \texttt{SALTshaker}}
\label{sec:SALTshaker}
The \texttt{SALTshaker} code\footnote{Available at \url{https://github.com/djones1040/SALTShaker}}, first presented in \citetalias{Kenworthy2021SALT3}, was designed to allow the creation and training of SALT models in Python. Our primary goal here was to add an extra component. In order to extend the model with an additional component, as well as make future modification easier, we have refactored the code to use the \texttt{JAX} library, which compiles Python code to optimize performance and allow automatic differentiation of arbitrary functions via the chain rule \citep{jax2018github}. The \texttt{SALTshaker} code uses the derivatives of the likelihood function to guide the optimizer; previously, these derivatives were coded by hand.  By relying on the autodiff functionality of the \texttt{JAX} library we can now allow the inclusion of arbitrary, user-defined functions for constraints, priors, and color laws without requiring a user to make multiple modifications across the program structure. As a result, the new code is more performant, modular, and expandable.

Concurrently the optimization process has been simplified. The Gauss-Newton optimizer included in the original code has been retained, but a new optimizer using a gradient-descent method has been used in this work. The original code alternated optimization of the flux and error model during the core fitting loop. Our revised optimizer uses a short burn-in period wherein the error model parameters are fixed while the flux model is fit before ``turning on'' the error model and fitting these parameters simultaneously. However we find that allowing the color scatter to be fit simultaneously with the rest of the model results in undesirable behavior due to the regularization prescriptions. Evaluation of the color scatter takes place after optimization has otherwise terminated, and only the color and color law parameters are allowed to be free during this step. Testing has shown that differences between the final surfaces trained with the new and old optimizer are present only at the level of mmag. The upgrades to the codebase have greatly improved the speed of the code from ~1 day for a full training on the original \citetalias{Kenworthy2021SALT3} sample on a laptop, to $\sim 2$ hours \textit{using the expanded sample} with the revised code. 

\subsection{Model description}
\label{subsec:model}
We here introduce \modelname, a new model which extends the dimensionality of the SALT framework to model more complex light-curve features. 

We model the flux of a SN\,Ia as a function of phase and wavelength

\begin{align} \label{eq:spectralfluxmodel}
    F(p,\lambda) =& \textrm{max}(0\ ,\ x_0 [M_0(p,\lambda;\boldsymbol{m_0}) + x_1 M_1(p,\lambda;\boldsymbol{m_1})\nonumber\\ +& x_2 M_2(p,\lambda;\boldsymbol{m_2})]   \cdot \exp(-0.4 
    \cdot c \cdot CL(\lambda;\boldsymbol{cl}))) ,
\end{align}
where $x_0$ represents the overall flux normalization, $\{ x_1, x_2\}$ model the intrinsic variation of the SNe\,Ia, and $c$ a color parameter accounting for intrinsic and/or extrinsic dust reddening\footnote{There is an error in Eq. 1 of \citetalias{Kenworthy2021SALT3}, which neglected the factor of $-0.4$ in the exponential term.}. Flux surfaces $\{M_0,M_1,M_2\}$ are spectral energy distributions (SEDs) defined on a basis of two dimensional, third-order B-splines. $CL(\lambda;\boldsymbol{cl})$ is a continuous piecewise function (with continuous first derivatives) defined as a polynomial between $2800$ and $8000$ \AA, and linear in wavelengths outside of that range. The quantities $\{\boldsymbol{m_0}, \boldsymbol{m_1}, \boldsymbol{m_2}, \boldsymbol{cl} \}$ are parameters controlling these elements of the model, which are determined by the training code. In addition to including an additional component, we explicitly required the flux to be positive. The flux equation is integrated over each bandpass, to determine photometric fluxes. In the training, spectroscopic fluxes are further modified by a recalibration factor of $\exp(\sum^{N_\textrm{Recal.}}_{i=0} a_i \lambda^i)$, where $a_i$ are polynomial coefficients, fitted for each spectrum. The $N_\textrm{Recal.}$, the number of spectral recalibration coefficients per spectrum are set as configuration options.  This factor ensures that the calibration of the model is sourced from photometry, rather than spectra, while retaining spectral information about local features.

Diversity in the SN\,Ia population that is not described by the flux surfaces $\{M_0,M_1,M_2\}$ is accounted for by a two term variance model. The first is the ``error model'', given as a function of phase and central filter wavelength for a given photometric band $\lambda_c$

\begin{align}
    \mathbf{\mathit{\Sigma}}=& \begin{pmatrix}
\sigma_{M_0,M_0}(p,\lambda_c ) & \sigma_{M_0,M_1}(p,\lambda_c ) & \sigma_{M_0,M_2}(p,\lambda_c )\\
\sigma_{M_0,M_1}(p,\lambda_c ) & \sigma_{M_1,M_1}(p,\lambda_c ) & \sigma_{M_1,M_2}(p,\lambda_c ) \\
\sigma_{M_0,M_2}(p,\lambda_c ) & \sigma_{M_1,M_2}(p,\lambda_c ) & \sigma_{M_2,M_2}(p,\lambda_c )
\end{pmatrix} \nonumber \\
    \mathbf{\mathit{x}} =& \begin{pmatrix}
        1 \\ x_1 \\ x_2
    \end{pmatrix} \nonumber \\
    \sigma^2_f(p,\lambda_c)=& \left[ x_0 \exp(c \cdot CL(\lambda_c))  \right]^2 \mathbf{\mathit{x}}^T \mathbf{\mathit{\Sigma}}\ \mathbf{\mathit{x}} 
\end{align}
similarly to the SALT3 error model. Here, the error components $\sigma_{M,M}$ are each zeroth order B-splines, whose parameters are determined during training. These represent additional variability associated with each parameter, allowing the model to assign different uncertainties to different SNe. The second component is the ``color scatter'' $k(\lambda_c)$, a covariant uncertainty that allows light curves of the same SN in different bands to be coherently offset relative to one another. This component of the model is akin to chromatic models of intrinsic scatter like that of \citet{Guy2010} and the diagonal terms of the covariance matrix from \citet{Chotard2011}. The color scatter and total covariance matrix are defined 

\begin{align}
k(\lambda;\boldsymbol{a})= & \exp (\sum^4_{i=0} \boldsymbol{a}_i \lambda^i)\\
    (\Sigma_{\text{Model}}) _{ij} = &\delta_{ij} \sigma^2_f(p_i,\lambda_{c (i)}) \nonumber \\+  &\begin{cases} 
        k^2([\lambda_c]) (\Vec{f}_\text{Model})_i (\Vec{f}_\text{Model})_j & X_i = X_j\\
        0 & \text{otherwise}
    \label{eq:photcovariance}
\end{cases}
\end{align}

where $\delta_{ij}$ is the Kronecker delta, $X_i$ is the photometric band in which the measurement was made, and $(\Vec{f}_\text{Model})_i$ is the predicted flux for a given data point. The parameters controlling the components, color law, error model, and color scatter are determined during the training process by a log-likelihood minimization evaluated across the photometric and spectroscopic data (see \citetalias{Kenworthy2021SALT3} for the definitions of the likelihood terms). 
\begin{figure} 
    \centering
    \includegraphics[width=0.45\textwidth]{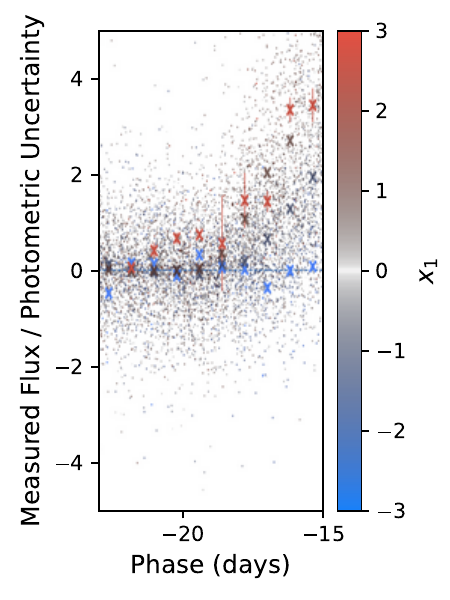}
    \caption{Early time photometric fluxes, relative to photometric uncertainties, in the expanded training sample. Each point is a single photometric measurement, colored according to the $x_1$ value of the supernova it belongs to. Crosses show mean values binned by both $x_1$ and phase. Negative flux measurements are possible due to measurement uncertainties in difference imaging/scene modelling. Observations of SNe with $x_1$ values close to the center of the distribution have been made partially transparent, to emphasize the edges of the population. PS1MD and Foundation data have been excluded from this plot, as some residuals showed indications of image subtraction issues.}
    \label{fig:detectresids}
\end{figure}

\subsection{Model Definitions and Priors}
\label{subsec:modeldef}
As the model described above is (mostly) linear, the fluxes modelled are invariant under a set of linear transformations among components and coordinates. As an example, a transformation which increases the magnitude of the components $M_0, M_1, M_2$ and decreases all $x_0$ by the same factor does not modify the final fluxes. To eliminate these degeneracies, we are required to make choices to uniquely determine the model. Previously, these definitions were enforced by narrow priors during training, then exactly enforced by post-processing after optimization concluded. The new code allows us to enforce  constraints exactly throughout the training, by transforming the parameters to satisfy the constraints at each likelihood evaluation, using \texttt{JAX}'s autodiff capabilities to differentiate through the transformation. Here we define the \modelname\ model to satisfy the conditions:

\begin{enumerate}
    \item { The rest-frame synthetic $B$-band flux of the $M_0$ component at peak is fixed such that $m_B^\text{peak}=10.5$ when $x_0=1$}
    \item { The rest-frame synthetic $B$-band flux of the $M_1$ and $M_2$ components at peak is defined to be 0 \footnote{By making this definition, the effects of $x_1$ and $x_2$ on the $B$-band maximum are absorbed into Tripp standardization parameters $\alpha_1$, $\alpha_2$.}
}
    \item The distributions of the light-curve parameters $c$, $x_1$, and $x_2$ over the training sample are defined to have 0 mean 
    \item The distributions of $x_1$ and $x_2$ have standard deviation 1
    \item The distributions of $x_1$, $x_2$, and $c$ have no correlation in the training sample
    \item The color law is defined such that $CL(4300 \textrm{ \AA}) = 0$ and $CL(5430 \textrm{ \AA})= -1$ , corresponding to central wavelengths for $B$ and $V$ bandpasses
    \item In post-processing after training is complete, the mutual information of the $\{ x_1, x_2\}$ distribution (as measured from a kernel-density estimator) is minimized by a rotation of the components and coordinates. We then label the component with larger RMS flux values as $M_1$ and the corresponding coordinate $x_1$, and the other pair as $\{M_2, x_2\}$. Lastly, the sign of both components is set such that they have positive $B$-band flux at 15 days. 
\end{enumerate}

The first 6 definitions are analogous to those made in SALT3, defining $x_0$ to represent the inferred peak $B$-band flux, $c$ to represent a $B-V$ color difference independent of any behavior with time-evolution, and $M_0$ to represent the SED of the mean SN\,Ia. The final definition is novel for SALT3+, although similar in concept to the fifth definition, and similar in principle to priors for variational autoencoders such as PARSNIP \citep{Boone2021ParSNIP}. The goal of the definition is to, as much as possible, separate different phenomena into different parameters; this makes later analysis easier.  
We find that the resulting $x_1$ parameter is strongly aligned with the previously derived SALT2/SALT3 $x_1$. Our modeling is entirely empirical and does not incorporate any theoretical information.

In a departure from the approach employed with previous SALT models, we include priors on the parameters during the training process. These priors are standard normals on the parameters $\{x_1,x_2\}$ as well as a normal distribution with mean 0 and standard deviation 0.2 on $c$. This approach was chosen because the second component $M_2$ is smaller in magnitude than $M_1$ and further is only significantly detectable at some points of the light-curve. As a result, for many SNe in the sample it is under-constrained, and the use of priors in the parameters helps to reduce the impact of overfitting in the training sample. As a further precaution against overfitting in the data, the surfaces are regularized by the use of penalty terms in the training objective function proportional to the derivatives; we use all three methods of regularization described in  \citetalias{Kenworthy2021SALT3}.

\subsection{Model Configuration}

In addition to the inclusion of an additional spectral surface, other changes have been made to the model configuration to ensure the model is well captured. The phase resolution of the underlying splines has been increased to 2.5 days from 3, while the wavelength resolution was kept identical to that of \citetalias{Kenworthy2021SALT3} at $69.3\ \AA$, sufficient to cover broad SN features.  All spectral recalibration polynomials were set to use a cubic polynomial for performance reasons, rather than setting the number of free parameters individually for each spectrum. 

\subsubsection{Early Time Data}

\begin{figure*}[t] 
    \centering
    \includegraphics[width=0.9\textwidth]{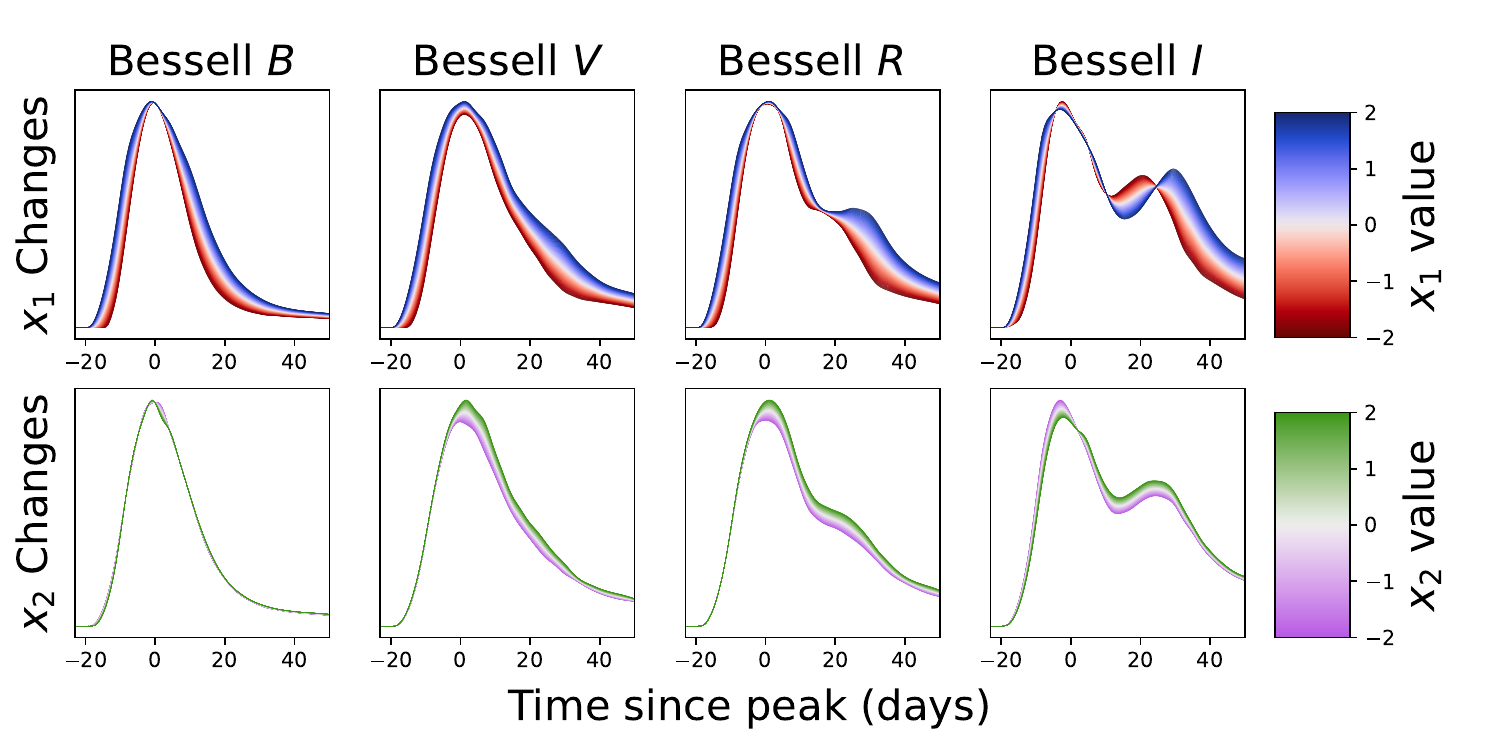}
    \caption{Light-curves for the \modelname\ model. Upper figure shows the variation in three bands as a function of $x_1$ with $x_2$ fixed to 0. Lower figure shows the variation as a function of $x_2$ with $x_1$ fixed to 0. Any other values will be represented by a linear combination of those light-curves.}
    \label{fig:multirange}
\end{figure*}

The SALT model requires the training code to specify a phase and wavelength range over which the model will be defined. Ideally, the model should begin immediately before the earliest SN explosions (relative to maximum light), with the $\{M_X\}$ components possessing no flux at that time and with the first derivatives zero. The third order B-splines can represent either an ``expanding fireball'' light-curve with $F\propto t^2$ and/or deviations through the cubic component of the splines.  Previous SALT models have used the phase range $-20$ days to $+50$ days. However, \citet{Miller2020ZTFRise} suggested, using early photometry of SNe\,Ia from the ZTF, that explosion may occur as much as 23 days prior to maximum light for some objects. 

For \texttt{SALTShaker}, ZTF data gives greatly enhanced phase resolution, doubling the number of photometric epochs (not necessarily detections) available between $-23$ and $-15$ days.  Thus, we examined whether the phase range needed to be expanded to better model the early-time evolution of the SNe Ia. In Fig. \ref{fig:detectresids}, we show early time photometric residuals relative to zero flux, roughly corresponding to detection significance. Only a small proportion of SNe can be individually detected at such early times, but in aggregate, we observe nonzero flux across the population as a whole. Visually, high $x_1$ SNe are detected earlier compared to low $x_1$ SNe, which can be undetectable even 15 days prior to  maximum light. 

Running \texttt{SALTShaker} with the previously standard phase range of $-20$ to $+50$ days, we noted that the flux in $M_1$ and $M_2$ surface at $-20$ days did not typically converge to 0; furthermore, adding a constraint on the model that the initial flux and/or the first derivative of the flux be 0 led to visually different model light-curves. Although no individual epochs are a detection at that early time, the aggregated weight of the entire sample is sufficient to affect the model. Running the model on this data with a minimum phase range of $-23$ days, we find that the resulting model, while showing small residuals at the edges of the phase range, does not predict any SNe to have observable ($>1\sigma$) flux before $-21$ days. Therefore, our final model uses the phase range $-21$ days to $+50$ days, to ensure the entire early evolution of the population is captured without adding unused flexibility to the model.

\section{Results}
\label{sec:results}

\subsection{Model surfaces}

We first visualize the surfaces produced by the \texttt{SALTshaker} code, in order to describe the modeled behavior.

\subsubsection{Synthetic light-curves}

\begin{figure*}[t] 
    \centering
    \includegraphics[width=0.9\textwidth]{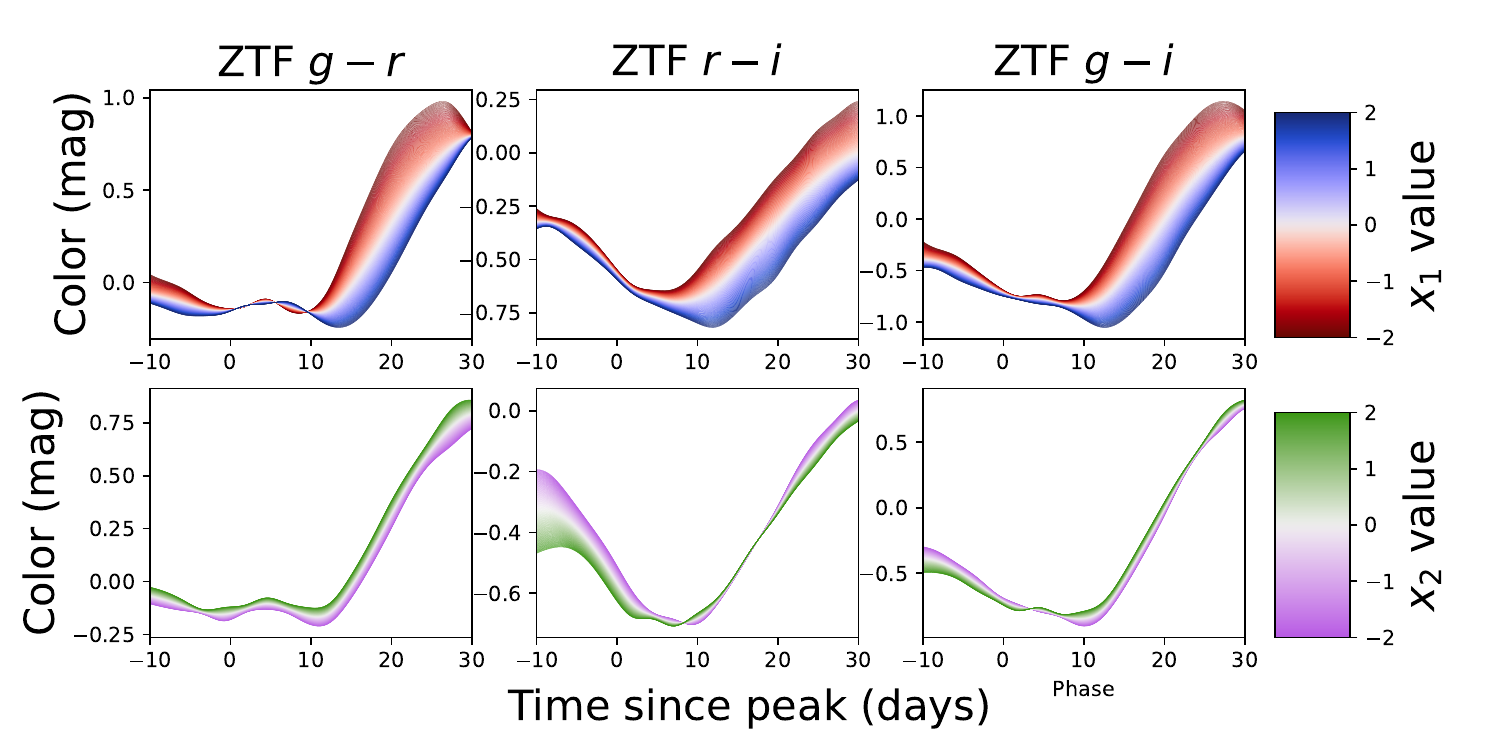}
    \caption{Color curves for the \modelname\ model in magnitudes. Upper figure shows the variation in three bands as a function of $x_1$ with $x_2$ fixed to 0. Lower figure shows the variation as a function of $x_2$ with $x_1$ fixed to 0. Any other values will be represented by a linear combination of those light-curves.}
    \label{fig:colorcurves}
\end{figure*}

As expected, adding a second component to the model is overall a relatively small effect. In Fig. \ref{fig:multirange}, we show synthetic light-curves over a range of parameter values. As can be seen, the size of the $M_2$ component is much smaller than the $M_1$ component. While $M_2$ is empirically derived over the whole phase and wavelength range, and is not associated in principle with any particular physical effect, we can qualitatively describe the major impacts on model light-curves. Increase in $x_2$ implies a higher peak and faster decline in $g$-band (as contrasted to $M_1$, which slows both the rise and decline times of the SN). In $i$ band, $x_2$ also increases the amplitude of the secondary maximum relative to the primary, while $x_1$ shifts the time at which the secondary maximum peaks. More extreme values of $x_2$ create a visible ``kink'' in the $i$-band light-curve in the primary peak. This feature was previously noted and discussed by \citet{Pessi2022Kinky}, who found that the strength of the feature showed little correlation with light-curve stretch, in agreement with our results here. $x_2$ is most measurable as a time-dependent effect in color curves, which are shown in Fig. \ref{fig:colorcurves}.  The contrast is most simply seen in $r-i$, where $x_1$ is visible as a constant spread around a fixed slope, while $x_2$ modifies the value of the slope.

\subsubsection{Spectral Effects}

\begin{figure*}[t] 
    \centering
    \includegraphics[width=0.9\textwidth]{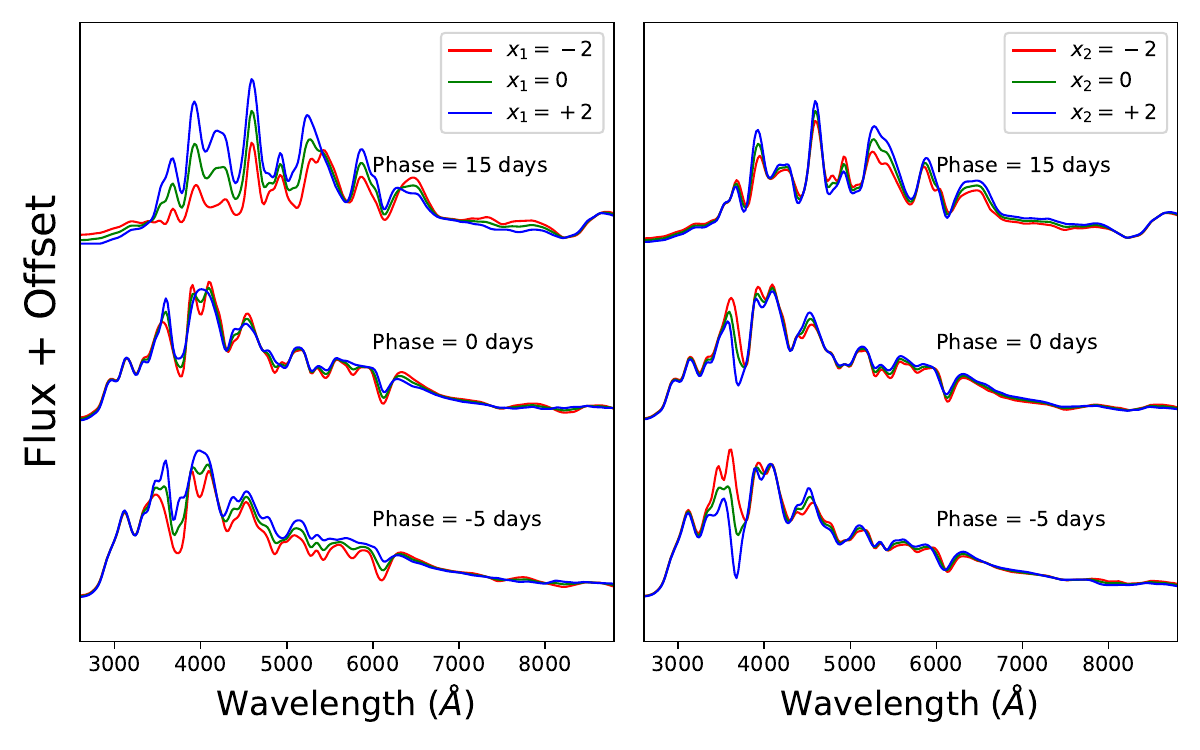}
    \caption{Synthetic spectra of \modelname\ across phase, $x_1$, and $x_2$. Left panel shows variation with $x_1$, right panel shows variation with $x_2$ from $-5$ days to $+35$ days.   }
    \label{fig:spectraldifferences}
\end{figure*}

\begin{figure*}[t] 
    \centering
    \includegraphics[width=0.9\textwidth]{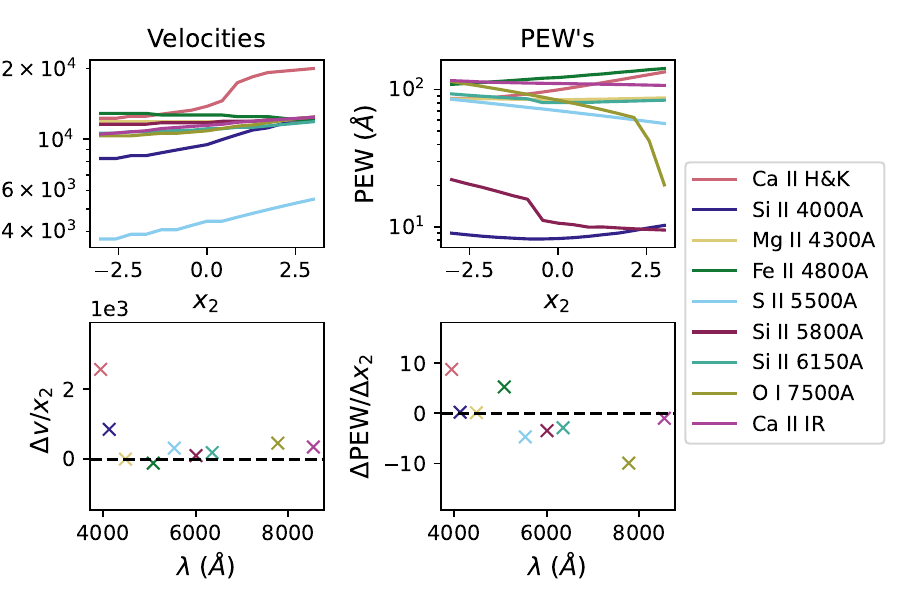}
    \caption{Spectral effect of $x_2$ at maximum light. Top left panel show velocity measured from spectral features as a function of $x_2$, while top right panel shows pseudo-equivalent widths (PEW) as a function of $x_2$. Bottom panels show slopes as a function of wavelength. }
    \label{fig:spectrends}
\end{figure*}

As SALT is a spectral model, besides examining light-curves we also inspect the effect of parameter changes on model spectra. Particularly at maximum light, the model has an abundance of spectral data to constrain the underlying surfaces. The spectra as a function of phase, $x_1$, and $x_2$ are plotted in Fig. \ref{fig:spectraldifferences}. To measure spectral features at maximum light, we use a fully automated and public code for spectral fitting, {\verb,spextractor,} \footnote{Code is publicly available at \url{github.com/astrobarn/spextractor}}  \citep{Papadogiannakis2019PhDT, Burrow2020}. The code uses a nonparametric, Gaussian process (GP) regression to get the minima for the individual features. It is based on the python package GPy \citep{gpy2014} and uses a Matern 3/2 kernel for smoothing the spectra. In Fig. \ref{fig:spectrends} we show how changes in $x_2$ affect the width and velocity of spectral features in model spectra at maximum light in B-band, where most SN spectra are taken. In general we see that increased $x_2$ is associated with a higher line velocity, with a wavelength-dependent effect on pseudo-equivalent widths. The iron line at $4800$~\AA\ is the only one to reverse the sign of the velocity trend.

Previous work has looked at the effect of spectral velocity on SN\,Ia light-curves and/or Hubble residuals. Our results indicate a  positive correlation between $g-r$/$B-V$ color and spectral velocity of $\sim 0.04$ mag, and a equivalently strong effect with opposite sign in redder (i.e. $r-i$) colors. Qualitatively, this is in agreement with previous work \citep{Wang2009SpectroscopicDistances,Mandel2014Colors,Dettman2021Velocity}. However, we do not presently see evidence of a host-galaxy correlation with $x_2$ (see Sec. \ref{sec:hosts}), while previous results have noted significant correlation between velocity and galaxy types \citep{Wang2013TwoPopulation,Pan2015HostVelocities,Pan2020Hostvelocity}. Velocity associations with mass were also not seen by the previous SALT reconstruction of \citet{Jones2023} or Burgaz et. al. 2024 using the ZTF sample. However $x_2$ likely does not capture all velocity variation in SN\,Ia, and/or improvements to the model may allow more precise measurements of host behavior.

\subsubsection{Model Uncertainties \& Color Dispersion}
\label{subsec:cdisp}

The \texttt{SALTshaker} code, in addition to estimating the fluxes characterizing the SN\,Ia population, also estimates the \textit{unmodelled} variation. This includes a component that is assumed to be stochastic, and will thus typically contribute little to the final error budgets. However a much more significant factor is the color dispersion, $k(\lambda_c)$. This gives a relative, covariant uncertainty on each individual band in a light-curve. As a consequence of the covariance, the color dispersion, for any SALT model, sets a fundamental error floor on the parameters $c$ and $m_B$ which cannot be reduced by additional epochs/more precise photometry, only additional filters. Further, the color dispersion is the only means provided to the model to account for phenomenology too complex to be included in the parametrization. As a result, the size of the color dispersion represents both a \emph{minimum} statistical uncertainty, and an indicator of the size of potential systematic uncertainties in a given model.

\begin{figure}
    \centering
    \includegraphics{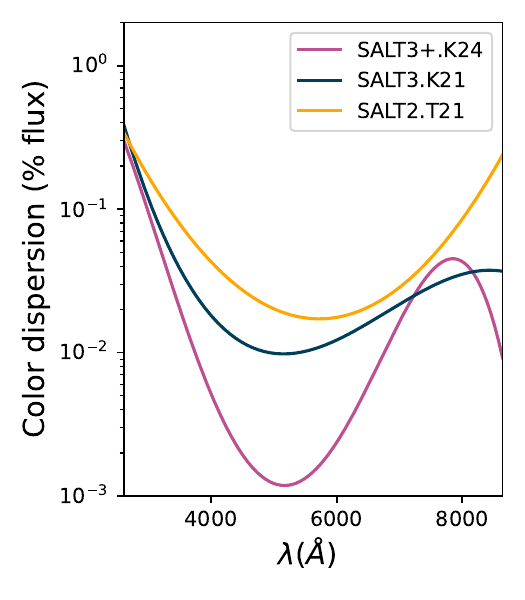}
    \caption{Color dispersion, defined in terms of \% uncertainty in the overall normalization of a given photometric band for a given supernova. Wavelength range shown here is the range of central filter wavelengths present in the training sample.}
    \label{fig:colordispersion}
\end{figure}

Accounting for $x_2$ results in a significant reduction in the color dispersion. We compare the color dispersion of SALT3.K21 and SALT2.T21 to \modelname\ in Fig. \ref{fig:colordispersion}. As can be seen, the color dispersion is substantively lower for our version of the model across most wavelengths used in cosmological analysis, with the exception of the $i$ band.  In the rest frame $V$-band, color dispersions are reduced to the mmag level, well below typical photometric uncertainties. Consistently high color scatters in the $i$ band may indicate difficulties with correctly fitting dust across all bands simultaneously with only a single color parameter. Studies such as \citet{Amanullah2015DiversityExtinction}, \citet{Mandel2017}, \citet{Brout2020}, \citet{Johansson2021NIRDust}, and \citet{Grayling2024BayeSNScalable} indicate that SN Ia colors are likely sourced from both extrinsic and intrinsic sources, which may contribute to differential wavelength dependence across the SN\,Ia population. Variance in dust population parameters such as $R_V$ from galaxy to galaxy may be a significant effect (c.f.\ \citealp{Popovic2021Dust2Dust,Wojtak2023TwoPop,Grayling2024BayeSNScalable}). Preliminary cross-calibration studies (\textit{Dovekie}, Popovic et. al. in prep) indicate an improved calibration solution for ZTF leads to a color scatter below that of SALT3.K21 in $i$ band, although still a local maximum. Other causes of systematic uncertainty in the $i$-band may include internal calibration of different surveys, calcium features, or further physics  beyond $x_2$ associated with the second peak.

\subsection{Light-curve fitting}
\label{subsec:lcfitting}
We fit our model to light-curves to the ZTF sample of light-curves, as well as the K21 compilation. We apply less restrictive light-curve quality cuts more typical of cosmology analysis than required for the training sample.

\begin{itemize}
    \item At least one measurement in at least two bands after peak brightness ($5 < p < 20$), to constrain the shape and color.
    \item At least one {\diff epoch} between $-20 < p <-1$ to ensure a constraint on time of maximum.
    \item At least six epochs in any bands between $-10 < p < 35$.
    \item Classified as a normal SN\,Ia, or belonging to the 91T subtype.
\end{itemize}

We perform our light-curve fits using \texttt{SNANA} \citep{Kessler2009SNANA:Analysis}, which has been upgraded to allow multi-component SALT models to be fit. Our fit is maximum likelihood based, although with we include unit normal priors centered at 0 on $x_1$ and $x_2$.

\subsubsection{Parameter distributions}

\begin{figure}   
    \centering
    \includegraphics{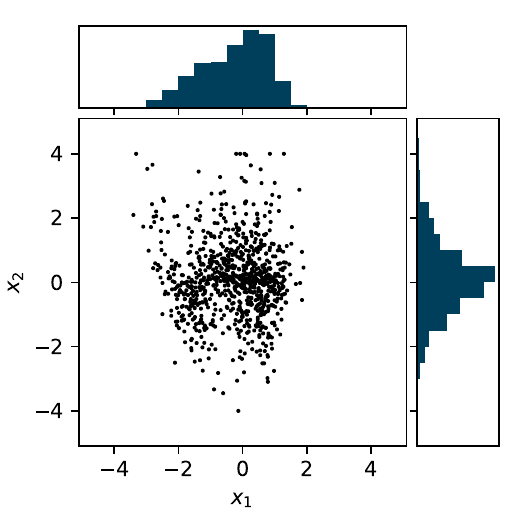}
    \caption{Distribution in two dimensions of the $x_1$,$x_2$ parameter space for the ZTF sample using only SNe Ia within the volume-limited sample ($z<0.06$)}
    \label{fig:paramdist}
\end{figure}

The shape of the parameter distributions could contain information about the underlying physics of the SN Ia populations. While the scale and location of the distribution are set by the model definitions (see Sec. \ref{subsec:modeldef}) and are thus not informative, the other moments of the distribution may provide information (see for example Ginolin et al. 2024). We show the distribution of parameters $x_1$,$x_2$ fitted to the sample in Fig. \ref{fig:paramdist}. The shape of the $x_1$ distribution is familiar relative to that found in other work (see also Ginolin et al. 2024 from the ZTF DR2; \citealp{KesslerScolnic2017,Nicolas_2021,Wojtak2023TwoPop}), showing a negative skew. The shape of the $x_2$ distribution is far more Gaussian. Understanding the intrinsic shape of the $x_2$ distribution however will require deconvolution with the noise, which forms a substantial portion of the observed distributions due to the high $x_2$ uncertainties.

\subsubsection{Parameter uncertainties}

\begin{figure*}[t]
    \centering
    \includegraphics[width=0.9\textwidth]{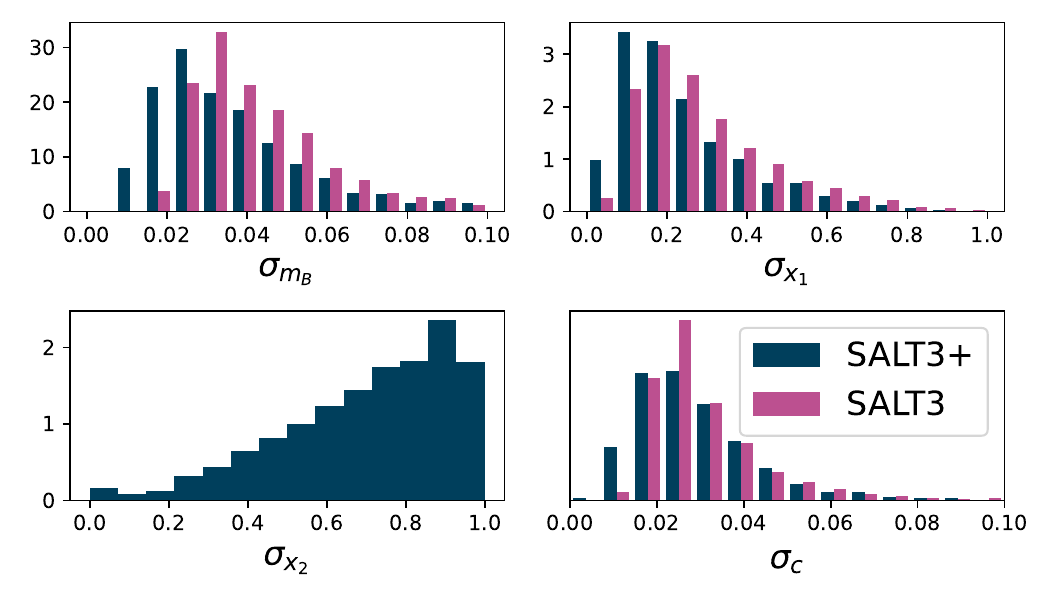}
    \caption{Distribution of parameter uncertainties as calculated under each \modelname\ and SALT3.K21.}
    \label{fig:uncertaintydist}
\end{figure*}

We show the estimated parameter uncertainties in Fig. \ref{fig:uncertaintydist} for each of the parameters. As discussed in Sec. \ref{subsec:cdisp}, the color dispersion bounds from below the uncertainties in color and magnitude, resulting in very few objects with color uncertainties <0.02 mag for SALT3.K21. The reduction in color dispersion effectively removes this floor, eliminating the sharp edge in the distribution. Magnitude and $x_1$ uncertainties are reduced across the board. However, $x_2$ is poorly constrained for most objects, and is dominated by the $N(0,1)$ prior.

\begin{figure}[t]
    \centering
    \includegraphics{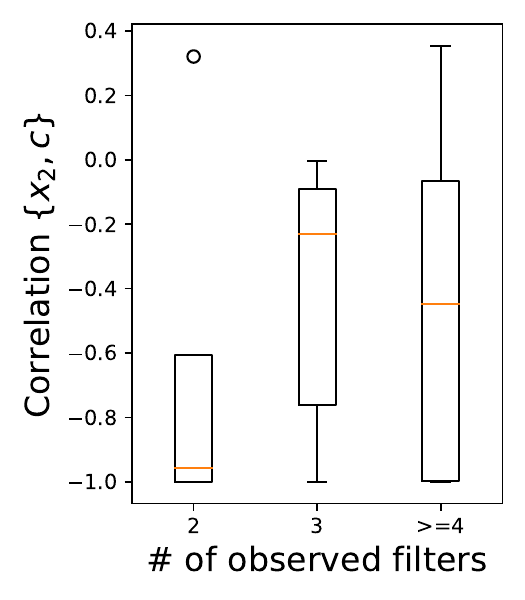}
    \caption{Boxplot showing the distribution of reduced correlation between $c$ and $x_2$ uncertainties, grouped by the number of filters observed, among SNe from surveys other than ZTF.  Correlation coefficient can range between -1 and +1.}
    \label{fig:x2corrbyfilters}
\end{figure}

\begin{figure}[t]
    \centering
    \includegraphics{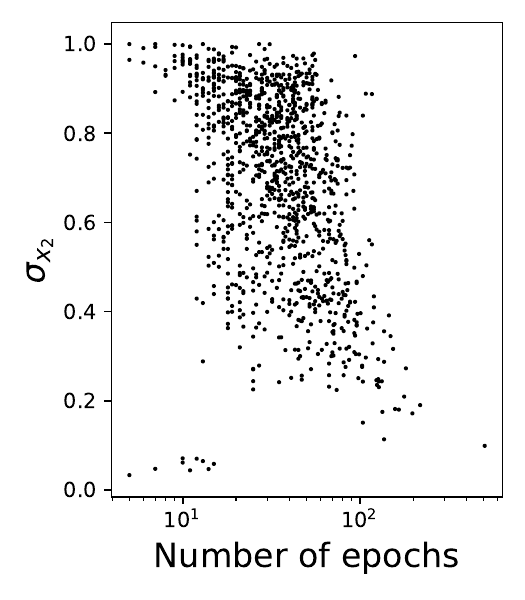}
    \caption{Distribution of uncertainty in $x_2$ as a function of the number of epochs in the fit. }
    \label{fig:x2errorvsepochs}
\end{figure}

We investigate some of the factors that contribute to an effective measurement of $x_2$. Fig. \ref{fig:x2corrbyfilters} and \ref{fig:x2errorvsepochs} show that SNe observed with more filters reduce the correlation between $x_2$ and $c$, and that uncertainty in $x_2$ shows dependence on the total number of measured photometric epochs.  In particular, at least three photometric filters are typically necessary to separate $c$ from the $x_2$ parameter to any significant degree.

\subsubsection{Comparison of SALT3.K21 and \modelname}

\begin{figure}
        \centering
    \includegraphics{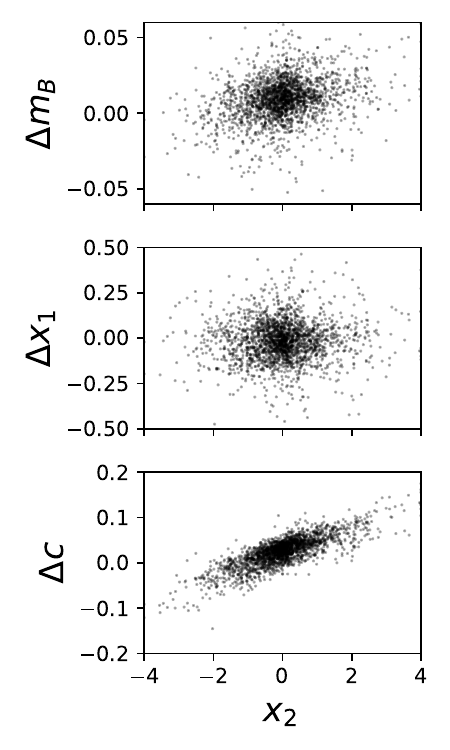}
    \caption{Difference in fitted light-curve parameters between SALT3.K21 and \modelname\ as a function of fitted $x_2$. }
    \label{fig:paramchanges}

\end{figure}

Changes in light-curve model affect ultimate cosmology results through changes to the fitted parameters. In Fig. \ref{fig:paramchanges} we show the difference in SALT3.K21 parameters relative to \modelname\ parameters, as a function of fitted $x_2$. A linear trend in $\Delta c$ is evident by eye and highly correlated, with Pearson $r=-0.84$, and a slope of $0.0251 \pm 0.0006$. The $x_1$ difference $\Delta x_1$ shows an effect of marginal significance, with a slope of $0.0062 \pm 0.0025$. We conclude that fits with SALT3.K21, ignorant of the $x_2$ effect, attempt to compensate with the two ingredients available to the fitter, $x_1$ and $c$. These trends indicate that while $x_2$ is a significantly smaller effect than $x_1$, and often has large uncertainties, fitting SNe\,Ia \emph{without} including $x_2$ in the fit will lead to significant changes in color parameters.

\subsection{Cosmology Implications}
\label{subsec:cosmo}
We create Hubble residuals using a simple linear standardization to allow comparison of cosmological performance between the previous SALT3.K21 model and the new \modelname\ model. In order to see the effect of the new model clearly, we cut the sample based on the fitted \modelname\ parameters with the following requirements:

\begin{enumerate}
\item The squared sum of $x_1$ and $x_2$ less than 9 (52 objects fail this cut)
\item $c$ between $-0.2$ and 0.4 mag (30 objects fail)
\item Estimated uncertainty in $c$ less than 0.1 mag (19 objects fail)
\item Estimated uncertainty in $x_1$ less than 0.5 (205 objects fail) 
\item P-value of fit greater than 1e-5 (185 objects fail)
\item Reduced correlation of $x_2$ and $c$ less than 0.9 (727 objects fail)
\end{enumerate}

Overall, of 2214 objects successfully fit by SNANA, 1161 (52\%) pass these cuts, of which the last cut is the most significant. For many objects, SNANA reports parameter uncertainties which indicate that the fitter is unable to constrain $x_2$ and $c$ independently. Many of these objects have been observed in only two filters; as $x_2$ is most significantly constrained by the effect on color-curves, this is problematic. In order to clearly see the effect of the introduction of $x_2$ on the cosmological nuisance parameters, we remove these SNe. Of the objects that pass other cuts, 120 do not have masses available from either Pantheon+ \citep{Brout2022PantheonPlus} or the ZTF DR2; for these, we include them in the fit and apply no mass step. {\diff Although we are fitting and validating on (mostly) overlapping samples, the Tripp estimator plays no role in the SALTshaker, and performance on these metrics is not evaluated in training. }

Given a SN at redshift $z$ in the CMB rest frame with fitted parameters ${x_0,x_1,x_2,c}$, associated covariance matrix $\Sigma$, and host galaxy stellar mass $M$, the Hubble residual $\Delta \mu$ is defined by analogue to the Tripp estimator \citep{Tripp1998b}

\begin{align}
    \hat{\mu} = &10.5- 2.5 \log_{10}(x_0) + \alpha_1 x_1 +\alpha_2 x_2  \\ &  - \beta c - \gamma \theta(M ) - \mathcal{M \nonumber} \label{eq:hubbleresids} \\
    \theta (M)=&  \begin{cases} 
      0 & \textrm{If no mass available}\\
      1/2 & M> 10^{10} M_\odot \\
      -1/2 &  M <10^{10} M_\odot 
   \end{cases}
\\
    \Delta \mu = &\hat{\mu}  - \mu(z)
\end{align}

\noindent where $\mu(z)$ is the distance modulus calculated under flat $\Lambda$CDM with \citet{PlanckCollaboration2018} cosmological parameters. Corresponding Gaussian uncertainties are estimated as

\begin{align}
        \sigma_\mu^2=& \sigma_\textrm{int}^2+ \sigma_{\mu,z}^2 +\sigma_\textrm{lens}^2 + \Vec{x}^T \cdot \Sigma \cdot \Vec{x}\\
    \Vec{x}=& \begin{pmatrix}
        \frac{-2.5}{\ln(10) \cdot x_0 } &  \alpha_1 & \alpha_2 & -\beta
    \end{pmatrix}^T \nonumber 
\end{align}

\noindent where $\sigma_{\mu,z}$ is computed from a peculiar velocity uncertainty of  \SI{250}{\kilo\meter\per\second} using the derivatives of the distance modulus with redshift, and $\sigma_\textrm{lens}=0.055 z$ \citep{Jonsson2010Lensing}.

These Hubble residuals depend on nuisance parameters: $\mathcal{M}$ the overall normalization, intrinsic standardization coefficients $\alpha_X$, color standardization coefficient $\beta$ (roughly analogous to an ``average'' $R_B$ across the sample if colors are assumed to be dust reddening), ``mass step'' $\gamma$, and the unparameterized or ``intrinsic'' scatter in the Hubble diagram $\sigma_\textrm{int}$. We also find that the ZTF sample shows much higher intrinsic scatter than the data from the K21 compilation, so we allow the intrinsic scatter for ZTF to be independent of the intrinsic scatter of the other data. We compute these parameters for our sample by maximizing a Gaussian log-likelihood $\mathcal{L}= -\sum \log (\sigma_\mu) /2 + (\Delta \mu / \sigma_\mu)^2 /2 $, using light-curve fits made with both the new model and the most recent version of SALT3.K21 (setting $\alpha_2=0$ in the latter case). As an optimizer, we use Minuit \citep{James1975}, and report the parameter error estimates from the Hessian matrix derived. This simple approach will not correct for regression or selection biases, but allows us to demonstrate some of the differences between \modelname\ and SALT3.K21. We present the fitted nuisance parameters and likelihoods in Table \ref{tab:nuisancepars}.

\begin{table*}[]

    \centering
    \resizebox{\textwidth}{!}{%
\begin{tabular}{c|c|c|c|c|c|c|c|c}
\hline
Name & $\mathcal{L}$ & $M_0$ & $\alpha_1$ & $\alpha_2$ & $\beta$ & $\sigma_\textrm{int}^\textrm{nonZTF}$ & $\sigma_\textrm{int}^\textrm{ZTF}$ & $\gamma$ \\ \hline
Full Sample, SALT3 & 427.63 & -19.471(0.005) & 0.145(0.005) & - & 2.72(0.05) & 0.082(0.007) & 0.140(0.007) & 0.07(0.01) \\ 
Full Sample, SALT3+ & 461.85 & -19.413(0.005) & 0.111(0.005) & -0.019(0.005) & 2.94(0.05) & 0.088(0.006) & 0.135(0.006) & 0.07(0.01) \\ 
\hline
Non-ZTF, SALT3 & 289.21 & -19.450(0.006) & 0.143(0.007) & - & 2.74(0.07) & 0.078(0.007) & - & 0.04(0.01) \\ 
Non-ZTF, SALT3+ & 292.22 & -19.385(0.006) & 0.107(0.007) & -0.034(0.007) & 2.91(0.08) & 0.082(0.006) & - & 0.04(0.01) \\ 
\hline
ZTF only, SALT3 & 167.98 & -19.511(0.008) & 0.156(0.008) & - & 2.89(0.06) & - & 0.128(0.007) & 0.12(0.02) \\ 
ZTF only, SALT3+ & 214.26 & -19.457(0.007) & 0.121(0.007) & -0.009(0.006) & 3.20(0.07) & - & 0.118(0.006) & 0.11(0.02) \\ 
\hline
\end{tabular}

}

        \caption{Tripp nuisance parameters and maximum likelihood, compared between \modelname\ and SALT3.K21. Nuisance parameters are shown as fitted using either the full sample, only ZTF data, or only non-ZTF data.  Definitions are given in Equation \ref{eq:hubbleresids}. The full sample here includes 1161 SNe, of which 643 are ZTF objects and 518 are from other surveys.  }
    \label{tab:nuisancepars}
\end{table*}


The Tripp fits show significant evidence in favor of the higher dimensionality; the log-likelihood difference across the full sample of corresponds to a $\Delta \textrm{AIC} = 66.4$, favoring \modelname\ at $\sim 8.3 \sigma$ with the inclusion of a single extra parameter in the fit. Likelihood ratios favor \modelname\ in both ZTF and non-ZTF subsamples. RMS scatter in the data slightly decreases with the additional parameter, from 0.184 to 0.178 mag.  $\mathcal{M}$ differences between the models are not in general physically relevant, as the zero-point of the model is arbitrary, varying with mean properties of the training sample. Further, a blinding factor has been added to $\mathcal{M}$ for the ZTF sample, as discussed in Rigault et al. (2024). However we note that when splitting the data between ZTF and non-ZTF subsamples, significant differences are present in $\alpha_2, \gamma$, and $\beta$, which may indicate issues of generalizability from ZTF to other surveys. These may simply be indications of the different selection functions between surveys, or reflect the current incompleteness of the ZTF calibration solution. Additionally, while the extended model is preferred by both samples, the significance outside of ZTF is only $\sim 2.5\sigma$. 

\begin{figure}
    \centering
    \includegraphics{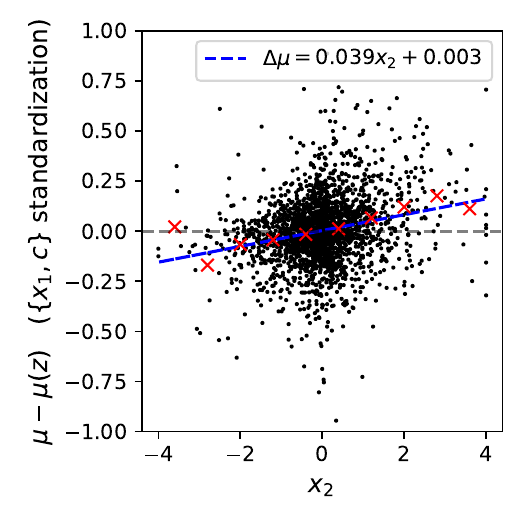}
    \caption{Hubble residuals calculated using SALT3.K21, with standardization in $x_1$ and $c$, against $x_2$ calculated using \modelname. Red crosses show binned residuals, with line of best-fit in blue.}
    \label{fig:hubblex2corr}
\end{figure}

Comparing the residuals between the models, the Hubble residuals made using the SALT3.K21 model show a trend in $x_2$ of $\sim 0.05$ mag which can be seen in Fig. \ref{fig:hubblex2corr}. The effect is primarily driven by the strong effect in the color parameter $c$ seen in Fig. \ref{fig:paramchanges},  with a small countervailing effect in $x_1$. If $x_2$ vary across populations, this could \emph{potentially} lead to a systematic of the same magnitude in one-dimensional SALT-based analyses. We examine our sample to see whether there is any evident trend of $x_2$ with redshift; for now, we do not see any apparent significant effect. Taking the difference in mean $x_2$ at $z<0.15$ and $z>0.15$, we find $\Delta x_2 = 0.054 \pm 0.036$, implying a potential systematic of $2.1 \pm 1.5 $ mmag. {\diff This systematic may increase under alternate prior assumptions; a full analysis would require a complete population model (further discussed in Sec \ref{sec:conclusions}).}

Interpreting differences in $\alpha$ between models is somewhat difficult. As the scales of $x_i$ are anchored by definition 4 in Sec. \ref{subsec:modeldef} to the demographics of the training sample,  $\alpha$ changes between models may reflect empirical differences or mere demographic changes in the model from including ZTF in the training sample. To determine which effect is driving the lowered $\alpha_1$ seen in \modelname, we need to anchor $\alpha_1$ to a specific light-curve feature. We use $\Delta m_{15}(B)$ since it was the first feature of the stretch effect discovered. Using synthetic light-curves, we evaluate $\nicefrac{\partial \Delta m_{15}(B)}{\partial x_1}$ for both models at $\{z=0,c=0,x_1=0, x_2=0\}$, then calculate $\nicefrac{d \hat{\mu}}{d \Delta m_{15}(B)} = \alpha \cdot  \nicefrac{\partial \Delta m_{15}(B)}{\partial x_1}^{-1} $ for both models; since the quantities present on the L.H.S. are light-curve features independent of sample demographics, this slope should be comparable between models (see also discussion in \citealp{Grayling2024BayeSNScalable}). We find $\nicefrac{d \hat{\mu}}{d \Delta m_{15}(B)}$ to be $0.796$ for SALT3 and $0.593$ for \modelname. This implies that the decrease in $\alpha_1$ seems to be a real difference between models, rather than a purely demographic difference. This may reflect increased scatter in measured $x_1$ values due to the additional degree of freedom in the fit. However, the increase in $\beta$ nevertheless allows SALT3+ to explain more scatter than the SALT3.K21 model.

\subsubsection{Colors}

\begin{figure}
    \centering
    \includegraphics{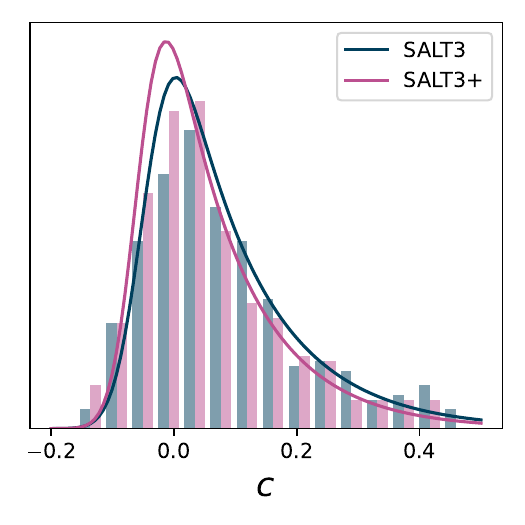}
    \caption{Observed and inferred intrinsic color distributions derived from the volume-limited ZTF sample. Solid lines show the convolved exponential distribution with parameters from Table \ref{tab:colorpars}. Histograms show the observed distribution, which are consistent with the inferred intrinsic distributions widened by noise. }
    \label{fig:colordists}
\end{figure}

As $c$ is defined by reference to a fixed quantity (that is, $B-V$ color), differences from model to model in $\beta$ across the same sample are more likely to be physically relevant. To evaluate the significance of the effect, we bootstrap resample the SNe and refit the Tripp estimator for both \modelname\ and SALT3.K21 with the resampled data. We estimate the difference between $\beta$ as $0.22 \pm 0.03$. The increase in $\beta$ seen between SALT3 and \modelname\ is likely here to be driven by increased accuracy in color measurements, bringing $\beta$ somewhat closer to $R_V$ values typically inferred from SED fitting of galaxy populations \citep[e.g.][]{Salim2018}. We note a difference in $\beta$ between ZTF and non-ZTF samples, which is stronger in the new model. This may be due to different dust demographics in the more complete sample of ZTF than previous data, explored in detail in \colorpaper.

\begin{table*}[]
    \centering
    \begin{tabular}{c|c|c|c|c}
    \hline
    Name & $\mathcal{L}$ & $\mu$ & $\sigma_{c,\textrm{int}}$ & $\tau$ \\ \hline
    SALT3 & 259.6 & -0.046 & 0.038 & 0.128 \\ 
    SALT3+ & 296.2 & -0.057 & 0.034 & 0.115 \\ 
    Bootstrapped Diff. & 36.5(5.9) & -0.011(0.005) & -0.005(0.006) & -0.012(0.004) \\ 
    \hline
    \end{tabular}
        \caption{Intrinsic color distribution parameters, compared between SALT3 and SALT3+. Resampled differences between the models are shown in the third line, with means and uncertainties.}
    \label{tab:colorpars}
\end{table*}

We next look at the distribution of the color parameter. A standard approach in the field, also used in \colorpaper, is to assume that host galaxy dust shows an exponential distribution in optical depth $\tau$, while intrinsic color is normally distributed with mean $\mu$ and scatter $\sigma_{c,\textrm{int}}$. We use the ZTF supernovae to derive intrinsic distributions, assuming that the sample is volume limited below $z<0.06$. With fitted color parameter $c$ and associated uncertainty $\sigma_c$, the likelihood over the color distribution is

\begin{equation} 
    \mathcal{L}= \sum_{N_\textrm{SN}} \log E( c|\mu,\sqrt{\sigma_c^2 + \sigma^2_{c,\textrm{int}} },\tau) 
\end{equation}

using $E(\mu,\sigma,\tau)$ to denote the convolution of the PDFs of the exponential and normal distributions. We maximize this likelihood function over the 358 ZTF SNe passing all light-curve cuts, besides the cut on color, below $z<0.06$ and report the values  of the derived population parameters in Table \ref{tab:colorpars}. The SALT3+ colors show a narrower distribution in both $\sigma_{c,\textrm{int}}$ and $\tau$, with the latter difference proving more robust in bootstrap testing. The narrower distribution is consistent with the $x_2$ component accounting for a fraction of the observed color distribution.  We show the derived intrinsic distributions and observed distributions in Fig. \ref{fig:colordists}.

\begin{figure*}
    \centering
    \includegraphics{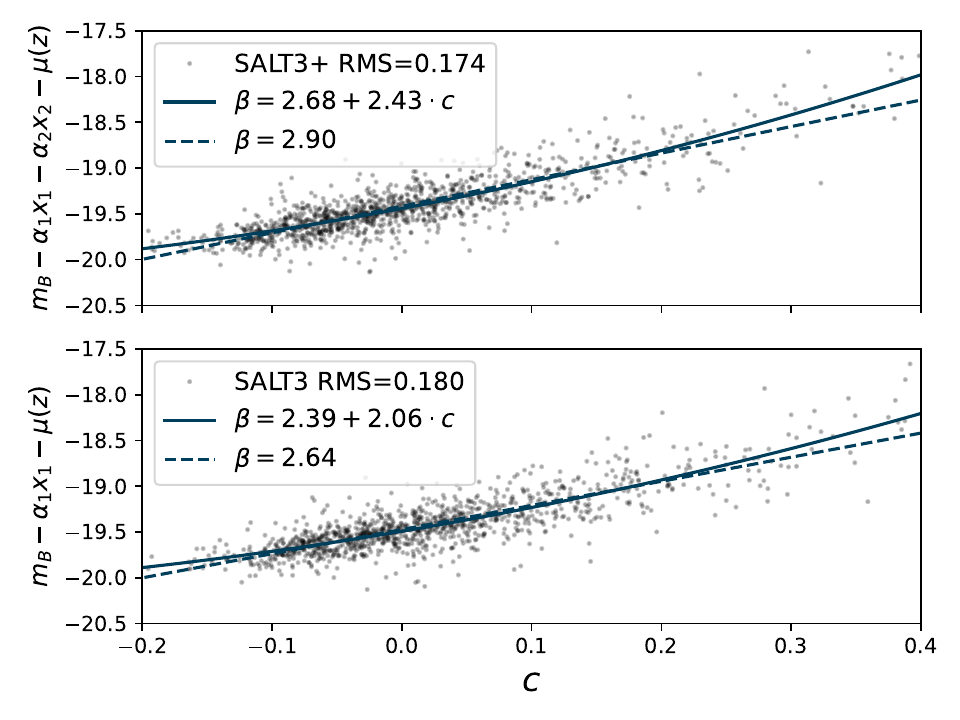}
    \caption{Hubble residuals, corrected for $x_1$ and $x_2$ but not color, as compared between models. Upper panel shows residuals calculated with \modelname, in lower panel with SALT3. Solid line shows a quadratic fit to the data, dashed line shows a linear fit. }
    \label{fig:colorresids}
\end{figure*}

In Fig. \ref{fig:colorresids}, we show the stretch-corrected, but not color-corrected, Hubble residuals. Previous analyses posited that red and blue SNe may obey different color laws \citep[see e.g.][]{Mandel2017}; current cosmology analyses such as \citet{Brout2022PantheonPlus} as well as \citet{Rubin2023UNITY} have explicitly or implicitly used a nonlinear color correction. This has been physically justified by appeal to distinct color-luminosity relations for host galaxy reddening in the red tail of the color distribution and intrinsic color in the blue tail. Here we note that the nonlinearity of the luminosity-color relation appears increased by the new model. To investigate further, we redefine the Tripp relation by setting $\beta$ in Equation \ref{eq:hubbleresids} to $\beta =\beta_0 + \beta' \cdot c$. Fitting the sample, a quadratic fit to the data is favored with either model, with a likelihood ratio of $\Delta \mathcal{L}=24.9$ for the full sample. With a single additional parameter, this corresponds to $\Delta\textrm{AIC}=47.8$. We then evaluate the difference in $\beta'$ between models by bootstrap resampling, finding $\Delta \beta'=0.35 \pm 0.16$, at $\sim 2.1 \sigma$ significance. This is consistent with an interpretation that the inclusion of $x_2$ in the fit improves the precision of $c$ as a tracer of host-galaxy extinction. Nonlinearity can also reflect regression dilution with an asymmetric underlying $c$ distribution (see the appendix of \colorpaper). The increased nonlinearity could be driven by reduced color uncertainties producing a $c$ distribution more strongly dominated by the (asymmetric) exponential tail. \colorpaper\ did not find strong evidence of an intrinsically nonlinear color relation in the ZTF sample using SALT2.T21-derived colors.

\subsubsection{Host Properties}
\label{sec:hosts}
\begin{figure}
    \centering
    \includegraphics{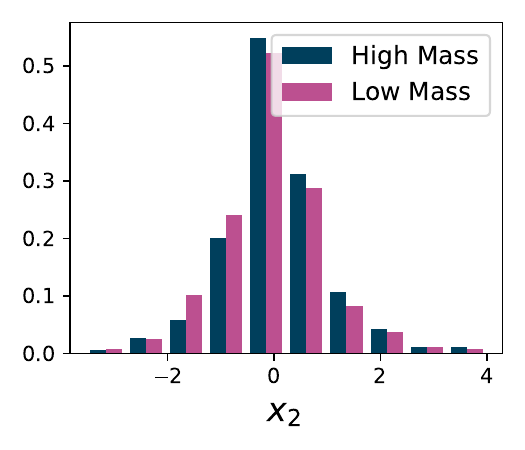}
    \caption{$x_2$ distribution, split between high and low mass galaxies at $10^{10} M_\odot$. }
    \label{fig:x2mass}
\end{figure}

We do not see any effective correlation between host galaxy properties and $x_2$. Likely, the significant noise present obscures any relation present. We show the distribution, split by host galaxy mass at $10^{10} M_\odot$, in Fig. \ref{fig:x2mass}. The mass step seems somewhat smaller in the ZTF sample, however this does not generalize to the non-ZTF sample.

\section{Conclusions}
\label{sec:conclusions}.
We present \modelname, a new SALT model with an additional principal component. Although preliminary, the model trained here has been added to both the \texttt{sncosmo} \citep{Barbady2015sncosmo,Barbady2016sncosmo} and \texttt{SNANA} \citep{Kessler2009SNANA:Analysis} codes for use in light-curve fitting. The new model shows particular improvement in producing informative colors for cosmology analysis.

However, the model as presented here is not ready for use in a full cosmology analysis. Our work here should thus be regarded as exploratory. The ZTF calibration is still a work in progress, with calibration uncertainties greater than centimag level. Accounting for these corrections can be forecasted to improve the uncertainties in the model (particularly $i$ band), and improve generalization from ZTF to non-ZTF surveys.  Further, we note that the issues seen with uncertainties in Sec. \ref{subsec:errors} and the additional intrinsic scatter in Sec. \ref{subsec:cosmo} may also affect the light-curve model. Future data releases from the ZTF collaboration plan to use scene-modelled photometry.  The Hubble residuals we presented have not been corrected for selection bias or regression dilution. The results on standardization presented here should thus be regarded as preliminary. Correctly accounting for these biases requires either frequentist forward-simulation, or a Bayesian hierarchical model. Frequentist analysis requires the derivation of parent populations suitable for use in a BBC-based analysis \citep{KesslerScolnic2017} through use of a code such as  Dust2Dust \citep{Popovic2021Dust2Dust}. Bayesian approaches include codes like UNITY \citep{Rubin2015UNITY, Rubin2023UNITY}, STEVE \citep{Hinton2019STEVE}, BAHAMAS \citep{March2011, Shariff2016, Rahman2022}, and others \citep[e.g.][]{Mandel2017, Feeney2018, Wojtak2023TwoPop} which can simultaneously derive cosmology and population parameters. Either approach will require further work. 

Possible physical origins of the effects we describe here are unclear. Spectral features indicating a velocity trend along with changes in broadband colors may reflect a different conversion of energy into thermal/kinetic components. Variation in velocity could indicate variation in the observed angle of an asymmetric explosion.  The significant changes in calcium features in our synthetic spectra could be a result of calcium plumes in the ejecta \citep{Khokhlov1995,Pessi2022Kinky}. Future work could investigate possible driving physics of these features.


{\diff As $x_2$ affects the rising light-curve less, our} results support the conclusions of \citet{Hayden2019StandardizationRise} that the most informative stretch information in the light-curves is in the rise of the SN. We also conclude that  SNe Ia must be observed in no fewer than three filters to robustly measure host galaxy extinction. Both of these emphasize the value of the ZTF data set in future cosmology. Further, evaluation of survey cadence strategies should likely incorporate these considerations. In addition, neglecting the second component in light-curve fits leads to a residual trend of $ 0.039 \pm 0.005$ mag in Hubble residuals; if there is demographic evolution in the second component, it could represent a systematic. However, we see no evidence for a shift in the mean of the $x_2$ distribution with redshift, leading to an estimated systematic of $\SI{2.1}{\milli\mag} \pm \SI{1.5}{\milli\mag}$. We conclude that any systematics from optical light-curve features of higher order are likely to be less than this value, and that models of dimensionality higher than this are unlikely to be a requirement. However, estimates of systematic uncertainties from forward simulation techniques such as BBC are sensitive to the model used for simulating data. Extended light-curve models may be important for these purposes. As $x_2$ shows little correlation with Hubble residuals as long as the phenomenology is included in light-curve fits,  it may be unnecessary to explicitly parametrize the phenomenon for use in cosmology analysis. The Gaussian process model for light-curve residuals used by BayeSN \citep{Mandel2020, Grayling2024BayeSNScalable} marginalizes over the light-curve variability, likely preventing an $x_2$ effect from affecting their parameters as seen in Fig. \ref{fig:paramchanges}. If accounting for $x_2$ at the level of cosmological analysis is unnecessary, developing a more robust error model for a one-dimensional SALT model may be a promising direction for future work.

\begin{acknowledgements}
The authors thank Daniel Kasen for useful discussion.

Based on observations obtained with the Samuel Oschin Telescope 48-inch and the 60-inch Telescope at the Palomar Observatory as part of the Zwicky Transient Facility project. ZTF is supported by the National Science Foundation under Grant No. AST-1440341 and a collaboration including Caltech, IPAC, the Weizmann Institute of Science, the Oskar Klein Center at Stockholm University, the University of Maryland, the University of Washington, Deutsches Elektronen-Synchrotron and Humboldt University, Los Alamos National Laboratories, the TANGO Consortium of Taiwan, the University of Wisconsin at Milwaukee, and Lawrence Berkeley National Laboratories. Operations are conducted by COO, IPAC, and UW. SED Machine is based upon work supported by the National Science Foundation under Grant No. 1106171

This work has been enabled by support from the research project grant ‘Understanding the Dynamic Universe’ funded by the Knut and Alice Wallenberg Foundation under Dnr KAW 2018.0067. AG acknowledges support from the Swedish Research Council, project 2020-03444. ST was supported by funding from the European Research Council (ERC) under the European Union's Horizon 2020 research and innovation programmes (grant agreement no. 101018897 CosmicExplorer). T.E.M.B. acknowledges financial support from the Spanish Ministerio de Ciencia e Innovaci\'on (MCIN), the Agencia Estatal de Investigaci\'on (AEI) 10.13039/501100011033, and the European Union Next Generation EU/PRTR funds under the 2021 Juan de la Cierva program FJC2021-047124-I and the PID2020-115253GA-I00 HOSTFLOWS project, from Centro Superior de Investigaciones Cient\'ificas (CSIC) under the PIE project 20215AT016, and the program Unidad de Excelencia Mar\'ia de Maeztu CEX2020-001058-M. JHT is supported by the H2020 European Research Council grant no. 758638. L.G. acknowledges financial support from AGAUR, CSIC, MCIN and AEI 10.13039/501100011033 under projects PID2020-115253GA-I00, PIE 20215AT016, CEX2020-001058-M, and 2021-SGR-01270. UB is supported by the H2020 European Research Council grant no. 758638 We acknowledge the University of Chicago's Research Computing Center for their support of this work. This project has received funding from the European Research Council (ERC) under the European Union's Horizon 2020 research and innovation programme (grant agreement n°759194 - USNAC). GD is supported by the H2020 European Research Council grant no. 758638. This project has received funding from the European Research Council (ERC) under the European Union's Horizon 2020 research and innovation programme (grant agreement n°759194 - USNAC). KM is supported by the H2020 European Research Council grant no. 758638. PN acknowledges support from the DOE under grant DE-AC02-05CH11231, Analytical Modeling for Extreme-Scale Computing Environments. Y.-L.K. has received funding from the Science and Technology Facilities Council [grant number ST/V000713/1]. 

\end{acknowledgements}

%
%

\bibliographystyle{aa}
\bibliography{references}
\end{document}